\documentclass[traditabstract]{aa}
\usepackage{graphicx,ulem,txfonts,amssymb}
\usepackage{aalongtable}
\input{epsf}
\def \xmm {{\it XMM-Newton\/}}

\def \h70 {$h_{70}$}
\begin{document}
\title{Merging history of three bimodal clusters \thanks{Based on data obtained with the 
European Southern Observatory, Chile (programs 072.A-0595, 075.A-0264, and 079.A-0425)}}
\author{ S.~Maurogordato\inst{1}, J.L.~Sauvageot\inst{2}, H.~Bourdin\inst{3},  
A.~Cappi\inst{1,4}, C.~Benoist\inst{1}, C.~Ferrari\inst{1}, G.~Mars \inst{1}, 
K.~Houairi\inst{5,6}}
\offprints{Sophie.Maurogordato@oca.eu}
\institute{
$^1$ Universit\'e de Nice Sophia-Antipolis, CNRS, 
Observatoire de la C\^ote d'Azur,  UMR 6202 CASSIOPEE, BP 4229, F-06304 Nice Cedex 4, France \\
$^2$C.E.A., DSM, Irfu, Service d'Astrophysique, C.E. Saclay,
  F91191, Gif sur Yvette Cedex, France\\
$^3$Dipartimento di Fisica, Universit\`a degli Studi di Roma "Tor
  Vergata", via della Ricerca Scientifica, 1, I-00133 Roma, Italy \\
$^4$ INAF - Osservatorio Astronomico di Bologna, via Ranzani 1, I-40127 
Bologna, Italy \\
$^5$ ONERA, Optics Department,BP 72, 92322 Chatillon Cedex, France \\
$^6$ CNES, 18 Avenue Edouard Belin, 31401 Toulouse Cedex 09, France
}
\date{Received date ; accepted date}
\abstract{
We present a combined X-ray and optical analysis of three bimodal galaxy 
clusters selected as merging candidates at $z \sim 0.1$. 
These targets are part of MUSIC (MUlti--Wavelength Sample of Interacting 
Clusters), which is
a general project designed to study the physics of merging 
clusters by means of multi-wavelength observations. Observations include 
spectro-imaging with XMM-Newton EPIC camera, multi-object spectroscopy 
(260 new redshifts), and 
wide-field imaging at the ESO 3.6m and 2.2m telescopes. We build 
a global picture of these clusters using  X-ray luminosity and temperature 
maps together with  galaxy density and velocity distributions. 
Idealized numerical 
simulations were used to constrain the merging scenario for each system. 
We show that A2933 is very likely an equal-mass advanced 
pre-merger $\sim 200$ Myr before the core collapse, while A2440 and A2384 are 
post-merger systems ($\sim 450$ Myr and $\sim 1.5$ Gyr after core collapse, 
respectively). 
In the case of A2384, we detect a spectacular filament of galaxies and gas 
%with low velocity dispersion ($\sim 200$ km/s) 
spreading over more than 
$1 h^{-1}$Mpc, which we infer to have been stripped during the previous 
collision.
The analysis of the MUSIC sample allows us to outline some general properties 
of merging clusters: a strong luminosity segregation of galaxies in recent 
post-mergers; the existence of preferential axes --corresponding to the 
merging directions-- along which the BCGs and structures on various scales 
are aligned; 
% corresponding to the merging direction.
the concomitance, in most major merger cases, of secondary merging or 
accretion events, with groups infalling onto the main cluster, and in some cases the evidence of previous merging episodes in one of the main components. 
These results are in good agreement with the hierarchical scenario of 
structure formation, 
in which clusters are expected to form by successive merging events, and 
matter is accreted along large--scale filaments.
}
\titlerunning{Merging history of three bimodal clusters}
\authorrunning{S.Maurogordato et al.}
\maketitle

\section{Introduction}

In standard Cold Dark Matter cosmological models, including the concordance 
$\Lambda$CDM, the general growth of structures starts from the 
primordial density fluctuations generated by inflation, 
and is driven by gravity in a hierarchical way, i.e. 
smaller structures form first, then merge into progressively more massive 
systems; however, if the
acceleration of the expansion is due to a cosmological constant, the process of structure formation will completely stop in the future  
(Krauss \& Starkman 2000; Nagamine \& Loeb 2003; Busha et al. 2005).

Galaxy clusters are the largest gravitationally bound objects of the hierarchy;
they accrete smaller groups coming from the filaments of the 
cosmic web and occasionally merge with other clusters of comparable mass, 
releasing an exceptionally high amount of energy.
Merging clusters are therefore ideal laboratories to study the
process of structure formation and how its affects galaxy evolution.
In this scenario, one expects to find a large fraction of clusters in their 
formation process at high redshift (which is corroborated by the 
large fraction of irregular morphologies observed). However, observations 
of clusters at high redshift are difficult and time consuming, so an 
alternative choice is to search for rarer but more easily observable 
merging candidates at low redshift. In this way, we can probe 
in detail the merging signatures and try to shed some light on 
the process of cluster formation and how it affects galaxy evolution.

A crucial progress in the study of merging clusters was provided by the
spectral and imaging capabilities of the last generation of 
X-ray satellites. Before then, X-ray spectroscopic information had not been available or of very poor quality, 
and the main information about mergers was based on morphology. 
With the spatially resolved spectroscopy and high resolution
imaging offered by Chandra and XMM, the situation has radically changed,  
and sophisticated algorithms have been developed to achieve good and reliable
 temperature maps (see for instance Bourdin et al. 2004), 
since the temperature is the more accurate tracer of the energy transfer from the collision to the X-ray gas itself. 
Strong signatures of the merging events have been detected in these 
maps (Vikhlinin et al. 2000, Markevitch et al. 2000 and 2002),
and cold fronts and bow shocks 
are now well established as common merger features.
Thanks to this observational progress, 
major mergers of galaxy clusters  now appear far more complex than 
previously foreseen. 

The full understanding of the complex processes at work in merging 
requires dedicated numerical simulations. 
Much progress has been made in 
this field, starting from the pioneering works of 
Schindler and B\"ohringer (1993) and Roettiger et al. (1997).
Ricker and Sarazin (2001) (hereafter RS01) described
the violent relaxation of gas in a dark matter potential well and
a variety of idealized merging systems, paying special 
attention to the impact parameter and the mass ratio between units. 
Poole et al. (2006) analyzed merging of idealized relaxed clusters with 
sophisticated SPH simulations including cooling and star formation, 
and detected the major transient signatures existing in 
observed temperature maps. 

Combining optical with X-ray data has been shown to be extremely 
effective in unveiling the complex history of merging clusters 
(to mention a few, Arnaud et al. 2000 and Maurogordato et al. 2000; 
Donnelly et al. 2001; Barrena et al. 2002; 2007; Boschin et al. 2004;
Owers et al. 2009). 
These studies have revealed various peculiar properties of the galaxy
distribution in the individual merging clusters, such as strong signatures
in the density and velocity distribution, and strong alignments effects. 
However, a larger sample with both X-ray and optical observations is 
clearly needed to test the generality of these properties, and their 
dependence on the merging stage.  

Motivated by these reasons, we started an observational program, 
MUSIC (MUltiwavelength Sample of Interacting Clusters) to define and analyze a sample of merging clusters at 
different stages of the merging process. The first targets 
(A3921, A1750, A2065, and A2255) were selected from 
the merging clusters observed during the XMM Guaranteed Time by one of us  
(J.L. Sauvageot), and we established an
optical follow-up program at ESO. 

To cover systems representing various stages of merging, 
we extended the initial sample by including targets 
from Kolokotronis et al. (2001), and compared the density distributions
of the gas (from ROSAT/HRI maps) and of the galaxies (from APM maps). 
We selected the systems with a clear bimodal morphology in both
the X-ray and the optical maps as probable pre-merger candidates, 
and those with a distorted morphology and a
pronounced segregation between gas (collisional) and galaxies 
(non collisional), i.e. with features that are
expected of mergers at a more advanced stage. We selected
clusters around $z\sim 0.1$, because at this redshift
evolutionary effects on galaxies are negligible and the field of view of 
X-ray and optical instruments offers a good coverage of the system 
(30' with XMM and wide field imagers such as WFI at the ESO 2.2m, 
corresponding to 2.3 h$^{-1}$ Mpc at $z=0.1$). With a 4m class telescope, spectroscopy is also easily performed as faint as
$R<19$, a limit corresponding to $\sim L_*+ 3.5$ at this redshift 
(with a best-fit parameter 
$M_R ^* = -22.97$ estimated by Popesso et al. (2005), fitting a Schechter 
function to the cluster galaxy luminosity function of the RASS-SDSS cluster 
survey). 
The aims of the program are to characterize the merging scenario (epoch of 
collision, geometry, and mass ratios) and to test the impact of the merging 
process on the properties of galaxies, in particular the star formation.
For example, the combined X--ray and 
optical analysis allowed us to determine the merging 
scenario of the first two clusters (a pre-merger for A1750,
Belsole et al. 2004, and a recent post-merger for A3921,
Belsole et al. 2005, Ferrari et al. 2005).

Here we present the analysis of three additional merging candidates: A2933, 
A2440, and A2384, based on  XMM X-ray spectro-imaging  obtained in 
2005 and 2006 (A2933 and A2440), or retrieved from the XSA database (A2384) 
and optical
observations carried out in 2003, 2005, and 2007 at ESO, including wide-field 
multi-band imaging with the Wide Field Imager on the 2.2m telescope, and 
multi-object spectroscopy with EFOSC2 at the 3.6m telescope. 

The clusters A2933, A2440 and A2384 share common features: an irregular morphology  
(they have been classified as  Bautz-Morgan Type III, II,
and II/III, respectively), low Abell richness classes (1, 0, and 1), 
and redshifts slightly less than 0.1 
(0.0925, 0.0906, and 0.0943 according to Struble and Rood 1999). 
In the optical, A2933 was observed as a target of the ESO Nearby Abell Cluster 
Survey (Katgert et al. 1996) and the southern Abell redshift 
survey (Muriel et al. 2002; Way et al. 2005; Coenda et al. 2006). 
A2440 was identified as a pre-merging cluster from the dynamical and 
X-ray/optical analysis of Beers et al. (1991) and Mohr et al. (1996).
A2384 is also a classical example of a  bimodal cluster 
(Ulmer and Cruddace 1982; West, Jones and Forman 1995). 
A weak-lensing analysis of A2384 was performed by Cypriano et al. (2004).
Both A2440 and A2384 belong to the flux-limited sample of bright clusters of 
galaxies from the southern part of the ROSAT All-Sky Survey 
(de Grandi et al 1999). 

In Sects. 2 and 3, we present the data, the reduction
procedure, and the methodology used in X-ray and optical respectively. In 
Sects. 4, 5, and 6, we analyze the data for A2933, A2440, and A2384 
respectively and propose a merging scenario for each cluster. In 
Sect. 7 we present our numerical simulations of merging clusters that
are used in Sect. 8 to refine the scenarios. 
In the following, we adopt the cosmological parameters of a
$\Lambda$CDM model with $\Omega_M$=0.3,
$\Omega_{\Lambda}$=0.7, and $H_0= 70\rm ~km ~s^{-1} ~Mpc^{-1}$. With
these parameters, at $z = 0.1$ one degree corresponds to a physical length 
of $6.6$ Mpc.

\begin{table*}[ht]
\caption{\xmm{}-EPIC observations used in our analysis, with 
effective exposure time corresponding to each instrument. 
In brackets: fraction of the useful exposure time after solar-flare
``cleaning''. \label{X_pointings_tab}}
\begin{center}
\begin{tabular}{cccccc}
\hline\hline
Cluster name & XMM-Newton  & Center coordinates & MOS1  effective  &  MOS2  effective &  PN  effective \\
& obs. IDs & & exposure time (ks) & exposure time (ks) & exposure time (ks)  \\
\hline
A2933 & 0305060101 &  01h40m41.2s -54$^\circ$33'26.0" & 21.5 (69.3  \%)  & 27.7  (89.0 \%) &  16.3  (68.5 \%) \\
A2440 & 0401920101 &  22h23m52.6s -01$^\circ$36'57.0" & 25.3 (55.4  \%)  & 24.9  (54.3 \%) &  17.8  (50.3 \%) \\
A2384 & 0101902701 &  21h52m14.2s -19$^\circ$42'19.8" & 17.2 (65.4 \%)   & 16.8  (63.6 \%) &  10.2  (51.6 \%) \\ 
\hline
\end{tabular}
\end{center}
\end{table*}

\section{X-ray observations: scientific products}

The three systems presented in this paper were observed by XMM-Newton EPIC
cameras for about 30 ks in full frame mode with the medium filter. The
observation were screened for proton flares on a high and low energy light
curve basis. A summary of the effective exposure time remaining after this screening
process is provided in Table \ref{X_pointings_tab}.

The brightness of the X-ray emitting intra-cluster medium (ICM) was
mapped from
a multi-scale algorithm using Haar wavelets, especially suited to denoise images 
dominated
by shot noise. First proposed by Jammal \& Bijaoui (2004), this algorithm was adapted to analyze
X-ray astronomical images and correct signal distorsions related to the spatially 
variable
detector response, and mirror effective area (Bourdin et al. 2008). We estimated the
ICM brightness from the spatial distribution of low energy X-ray photons 
(with energy lower than
2.5 keV), since their spectral distribution is expected to have a weak dependence 
on temperature.

We used the X-ray data set in order to map the ICM temperature.
To do so, we used the wavelet spectral-imaging algorithm proposed in Bourdin et al.
 (2004) and
Bourdin et al. (2008). X-ray spectral-imaging can be performed only on data of sufficiently high count statistics, 
and it is necessary to gather
photons within large detector regions, the shape of which needs to be optimised to detect thermal structure. To overcome this major difficulty, we developed
a strategy based on wavelet transforms, 
where we first estimated the ICM temperature with its
 associated confidence range at various angular scales and locations in the
field of view. The detection of the temperature features and
the reconstruction of an ICM temperature map was finally obtained by applying a
threshold to the wavelet coefficients. 
The wavelet analysis was performed following a
shift-invariant algorithm using B-spline wavelet coefficients, and applying a threshold of
$1 \sigma$ confidence level above the noise fluctuation. Point sources have
to be masked before building the map, to avoid pollution
from the ICM signal at different spatial frequencies.

\section{Optical observations: data description and reduction methods}

All three clusters were allocated time at ESO for wide-field imaging with
 the WFI instrument at the 2.2m telescope (total of 9.75 hours), 
and extensive multi-object spectroscopy with EFOSC2 at the 3.6m telescope 
(total of 6 nights), in 2003, 2005, and 2007, respectively 
(programs 072.A-0595, 075.A-0264, and 079.A-0425). 
As wide-field imaging of A2933 in service mode was not performed, we 
used the APM galaxy survey (Maddox et al. 1990) and the DSS images in the 
analysis of this cluster. We retrieved the relevant galaxy catalog in a 
region of 30'x30' covering A2933 from the Southern sky catalogue 
based on the UKST SES R survey. 
For A2384 and A2440, we obtained imaging of the central 30'x30' field 
in the R (filter ESO/844) and B (filter ESO/878) passbands. 
For each filter, 8 dithered images were obtained leading to a total exposure
time of 40min. These sets of images were reduced and combined using
the ESO/MVM package Alambic (Vandamme et al. 2002), and the catalogs
were extracted with SExtractor (Bertin and Arnouts 1996).
The typical seeing is 1 arcsec. 
The magnitudes used in the present paper are the total magnitudes 
"$MAG_{AUTO}$" provided by SExtractor. 
Stars and galaxies were classified (up to a magnitude
of 21 in both bands) by extracting the stellar sequences within the
magnitude--half light radius diagrams. Fainter objects were all
considered as galaxies. 
Finally, catalogs with $B-R$ color were built by associating objects with
positions in the $B$ and $R$ catalogs differing by less than
1.5 arcsec. 

For A2440 and A2384, the final catalog of galaxies including both $B$ and $R$ 
passbands and covering a field of 30'x 30' was used 
to analyze the color properties and the 
spatial distribution of the cluster galaxies. The color-magnitude diagram was 
used to identify the red sequence defined by the cluster elliptical population
 (Lopez-Cruz et al. 2004). We selected galaxies with colors 
within 3 $\sigma$ of this relation as high probability members of the cluster
 and from this selection we derived the projected density maps 
for different magnitude cuts in the R-band 
(corresponding typically to L*+1, L*+2, L*+3). In the case of A2933, 
R band magnitudes  from the APM catalog were used to compute density maps 
without color selection. 

\begin{figure*}
\centering
\includegraphics[scale=0.8,angle=0,keepaspectratio]{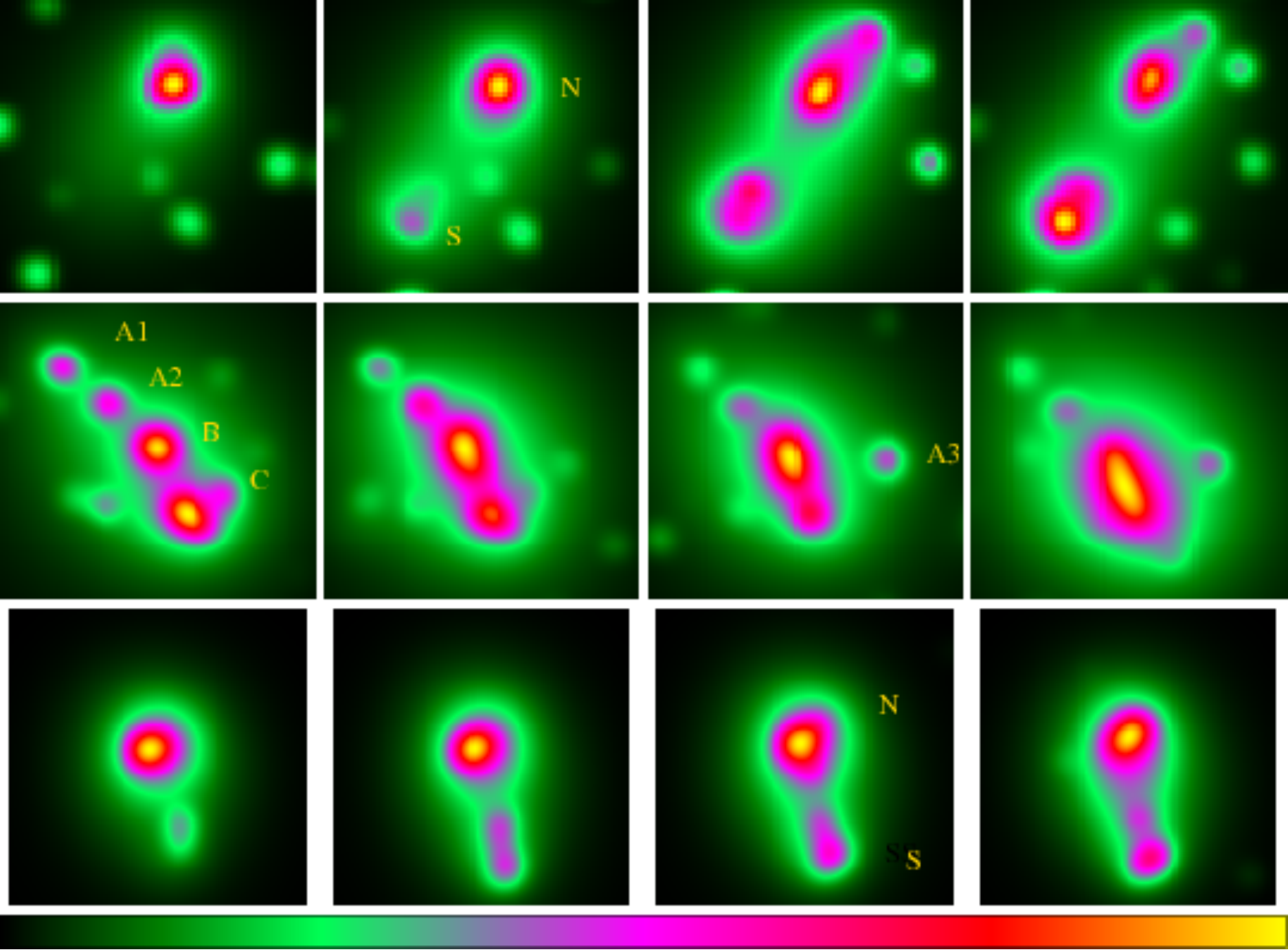}
\caption[]
{Galaxy density maps as a function of 
galaxy luminosity. From top to bottom: A2933, A2440 and A2384. The field of view displayed is 15'x 15' for A2933 and A2440, and 30'x30' for A2384. 
From left to right, magnitude cuts are: $R<18$, $ R<19$, $R<20$ and $R<21$.
The density maps of A2440 and A2384 have been computed for 
red sequence galaxies.}
\label{isodens_mag}
\end{figure*}

Spectroscopy was performed with EFOSC2 using the grism $\#03$, which 
covers the spectral range 3050-6010 $\AA$, with a FWHM of 7.5 $\AA$. 
In general, for each  pointing we obtained $2 \times 45$min 
exposures. Wavelength calibration was performed
in real time, taking spectra of arc calibration lamps (Helium-Argon) after
each exposure. Data reduction was performed using our dedicated IRAF package
{\it speXtra} for automatic extraction and wavelength calibration of 
spectra. Radial velocities were obtained using the
cross-correlation technique (Tonry and Davis 1981) with the {\it rvsao}
package. Cross-correlation was performed with velocity standards observed
with the same instrumental configuration during the observing run. 
%%% BEGIN ADD
The typical velocity error is $\sim 50$ km/s, an estimate
that we could confirm by performing a cross-check against the data available 
in the literature (see section 4.3).

%%% END ADD

The spectroscopic catalogs resulting from our observations 
are listed for the three clusters in 
Tables \ref{Table_A2933}, \ref{Table_A2440}, and \ref{Table_A2384}, where
columns are as follows:
1) identification number;
2) and 3) right ascension and declination (J2000.0);
4) radial velocity;
5) velocity error for the cross-correlation;
6) $R_{TR}$ parameter (Tonry and Davis 1981; when $R_{TR} > 3$, 
the cross-correlation redshift can be considered as reliable); and 
7) quality flag for the redshift (0: high precision, 1: medium precision).

Objects with a 
velocity within a $ \pm 5000$ km/s window 
centered on the mean velocity were selected
 as cluster member candidates. Possible interlopers were rejected by the 
gap technique. The velocity location and scale of the various clusters (and 
subclusters) were
determined with the biweight estimator using the ROSTAT package (Beers et al.
 1990). 

Ten normality tests (provided by ROSTAT) were
applied to the data. The Dip test for unimodality (Hartigan and Hartigan 1985)
was computed with the {\it diptest} package implemented by 
Martin Maechler in 
the R environment. The $P_{value}$ was computed in each case according 
to the value of the Dip test and the number of objects according to the table 
provided in Hartigan and Hartigan (1985).
The results of this analysis are displayed in 
Table \ref{table::rostat}, where columns are as follows: 
1) name of the cluster (or subcluster); 
2) and 3): right ascension and declination of the brightest galaxy (BCG) 
of the (sub)cluster; 
4) angular radius $\theta$ of the circle in which galaxies were
selected for the ROSTAT analysis; 
5) and 6) estimation of location and scale with the biweight technique; 
7) number of redshifts used in the analysis; 
8) radial velocity of the BCG;
9) number of statistical tests excluding Gaussianity 
at more than $10\%$ confidence; 
10) and 11) value of the Dip test and of its P-value; and
12) mean temperature computed within radius $\theta$ and its error. 

To test for multi-modality, we tried in each case to fit the entire velocity 
distribution  with a mixture of Gaussian functions. For this purpose, we have
used the powerful EMMIX algorithm (McLachlan and Krishnan, 1997; 
McLachlan et al.1999). This 
program is quite flexible and allows a variety of choices in the fitting. In 
particular, it does not require to introduce any guess for the initial partition
of velocities and the form of the covariance matrix. 
We applied EMMIX to the data, trying to fit the velocity distribution with 
mixtures of $N_g$ Gaussian distributions, with $N_g$ varying from 1 to 5.  
We retained the best fit on the basis of the P-value, 
and in the case of identical values of P-value, on the criterion of the smallest partition number. When a mixture of 
gaussians had been fitted to the data, we recomputed the value of the location and 
scale of each partition with ROSTAT.

%=============================================================================

\section{A2933}

\subsection{X-ray gas morphology and thermal structures}

\begin{figure*}[ht]
\centering
\hspace{-0.5cm}
\includegraphics[width=\textwidth]{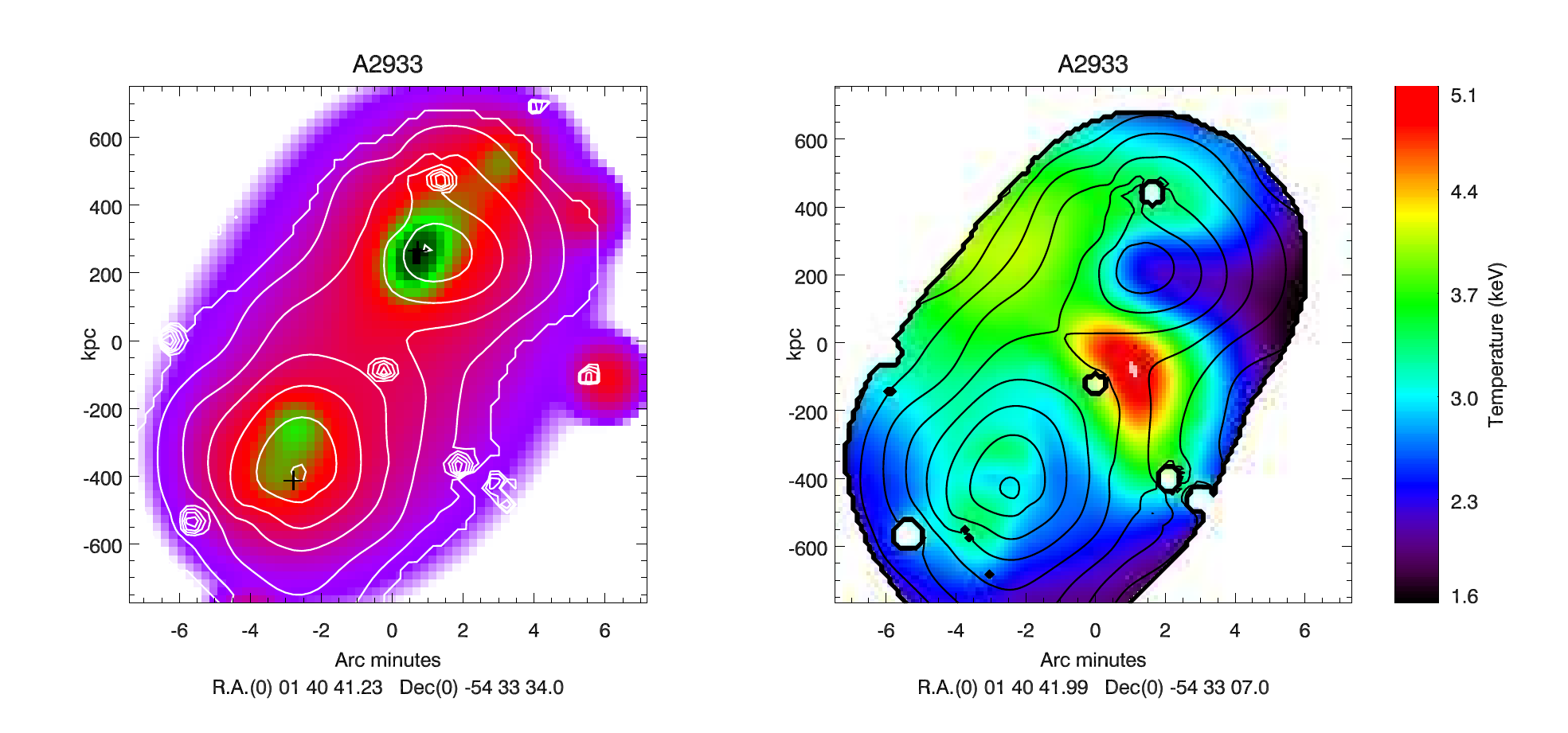}
\caption{Left: Galaxy density map of A2933 (magnitude cut: R$<$20) overlaid on 
the X-ray luminosity contours (EPIC-XMM counts in the 0.5-2.5 keV band 
corrected for background and vignetting). The black crosses indicate the 
BCG positions. Right: ICM temperature map of A2933 overlaid on the X-ray 
luminosity contours (EPIC-XMM data analyzed by means of wavelet spectral-imaging; 
see Bourdin et al. 2004, 2008 for details).}
\label{fig:OLT_a2933}
\end{figure*}

In X--rays, A2933 is a well defined bimodal system.
In Fig. \ref{fig:OLT_a2933} we show the temperature map 
of A2933 with the X-ray brightness contours superimposed.
Looking at the low energy contours, both components look quite 
regular and have similar brightnesses. We clearly detect the interaction 
zone as a hot region located in--between the two maxima.
The northern unit (A2933N) is slightly elongated along the east-west 
direction, while the southern unit (A2933S) is orientated along 
the general NW-SE direction connecting the two units. 
In the temperature map, the most prominent feature is the hot region 
between the two units, which is statistically significant,
because the exposure time of our XMM/EPIC observation was specifically adapted 
to ensure good statistics in the region of interaction.
The temperature distribution of A2933N appears to be quite elongated, 
while A2933S looks more regular.

\subsection{Galaxy density distribution}

The morphology of Abell 2933 has been very poorly studied at optical wavelengths.
In the following analysis, projected density maps 
were inferred in the field of the cluster
from galaxy positions and magnitudes in the APM catalog.

\begin{figure}
\centering
\includegraphics[angle=0,keepaspectratio,width=\columnwidth]{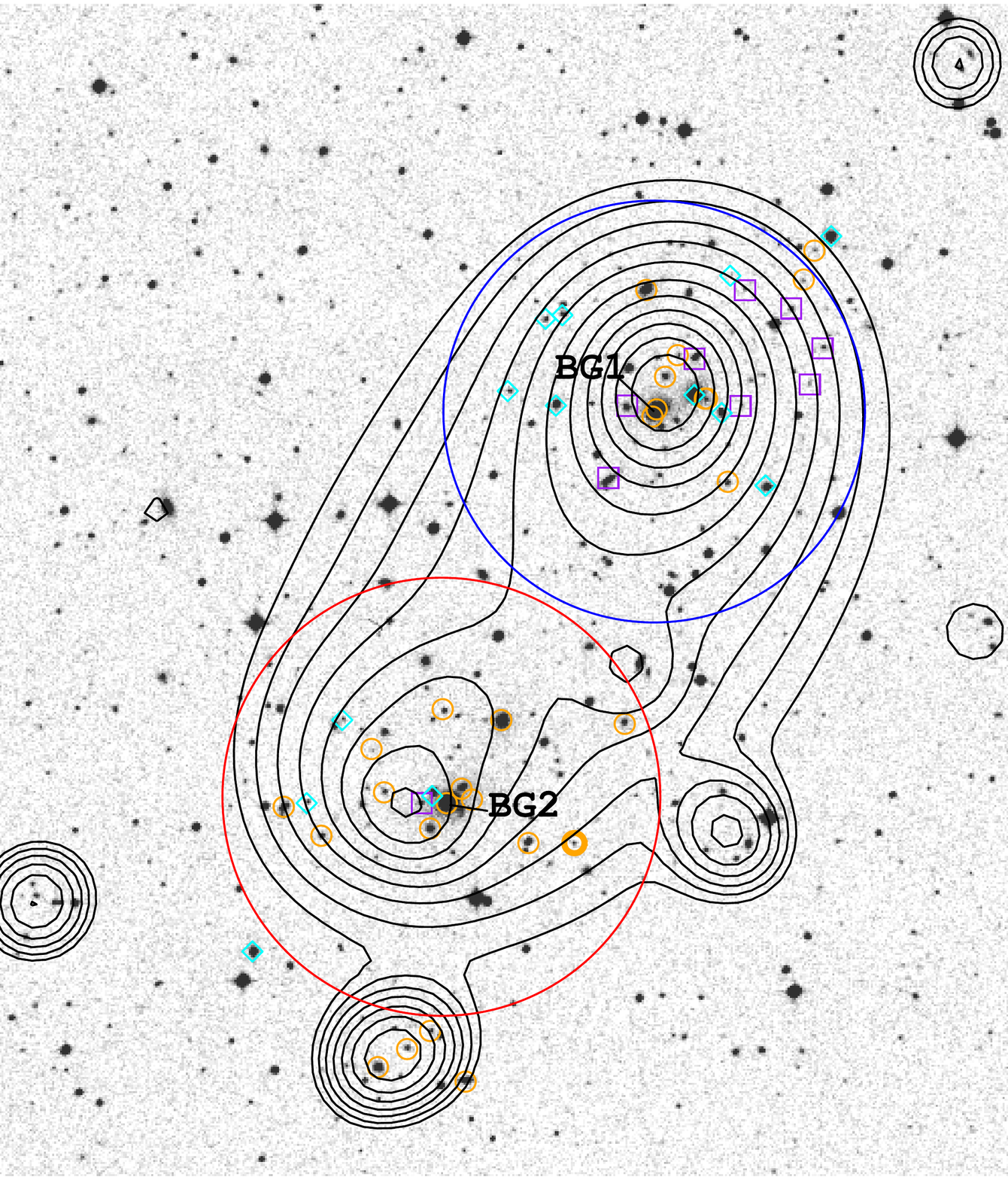}
\caption[]
{DSS region (17'x20') centered on A2933 (North is up and East is to the 
left). 
Galaxies identified as cluster members from spectroscopy are
marked with different symbols corresponding to the three major peaks 
in the velocity histogram (purple squares: [26000,26500]km/s, 
cyan diamonds:[26500,27200]km/s, 
orange circles:[27200,28500]km/s) which is plotted in 
Fig.\ref{fig:A2933_histo_emmix}. 
The isocontours corresponding to the galaxy density map with 
$R<19$ are superimposed. The velocity histograms for the galaxies in
the two circles (which have a 3.5 arcmin radius and are centered 
on the two subclusters) are displayed in Fig.\ref{fig:A2933_histo_subclus}.}
\label{A2933_spectro}
\end{figure}

The galaxy density distribution shows a 
bimodal structure, with two subclusters located on a NW/SE axis at a 
separation of $\sim 7$ arcmin (Fig.\ref{isodens_mag}). The relative importance
of the two components varies with the magnitude cut-off.
Selecting bright galaxies ($R<18$), the NW subcluster (A2933N) is dominant. 
When including fainter objects, the SE (A2933S) subcluster becomes
progressively more prominent (Fig.\ref{isodens_mag}). 
%and its centroid slightly shifts NW.
 At faint magnitudes, the NW subcluster shows an 
extended tail in the NW direction.
In Fig.\ref{A2933_spectro}, we also show
the isodensity contours corresponding to $R<20$, superimposed on the
DSS image of A2933. A2933N hosts a couple of very bright 
galaxies, and is centered on a BCG with a very close companion, while 
A2933S hosts a single bright elliptical galaxy. One can also note a small 
density clump at the southern extremity of A2933S. 
%At its Southern extremity, one can note 
%small clump of galaxies. 
There is a general alignment of the two BCGs, the two subclusters, and the 
global cluster structure along the NW/SE direction joining A2933N and A2933S.
%(PA=+38 and
%+51 deg for the Northern BCGs , PA=+33 deg for the Southern BCG, to be
%compared to PA=+49 deg for the NW subcluster and PA=+25 deg for
%the SE subcluster).

% FIGURE 2 OPTICAL/X-RAY MAPS OF A2933
 
\subsection{Galaxy velocity distribution}

A2933 was included in the Southern Abell Redshift Survey (SARS; Way et al. 
2005), and on the basis of 53 redshifts Muriel et al. (2002)
estimated the cluster global velocity and velocity dispersion 
($27\,709 \pm 105$ km/s and $759 \pm 72$ km/s, respectively); 
they also showed that the velocity dispersion as a
function of radius appears to be constant to 5 Mpc/h from the cluster center.
Their data, however, sample the cluster and its environment on large scales 
and at relatively bright magnitudes.
Since we are interested in the merging process,
we observed the central region including the two main clumps
at deeper magnitudes.

We measured 71 redshifts in the $30' \times 30'$ field of A2933, listed in 
Table \ref{Table_A2933}, from which 47 objects are identified as cluster 
members. 
We have only 11 redshifts (among 53) in common with them, because of the 
different region sampled. Excluding one discrepant case,
the redshifts of common objects are all in excellent agreement, 
%%% BEGIN ADD
with a mean velocity difference of 14 km/s and a standard deviation 
of 59 km/s.
%%% END ADD
From our redshift sample, we analyzed the velocity distribution in the 
central region of A2933. We measured a location of 
$C_{BI}= 27\,281 \pm 103 $ km/s and a scale of 
$S_{BI}= 682 \pm 65$ km/s (Table \ref{table::rostat}).
Our estimate of velocity location is significantly 
lower ($\sim 400$ km/s) than that found by
Muriel et al. (2002). These differences may be caused by variations 
in sampling.  Their sample covers a much larger area, that is not centered 
on the two subclusters but is shifted to the south, which we  show to be 
a region of higher mean velocity. We checked that their estimate of mean
velocity  and our estimate for southern component (see below) are consistent. 

\begin{figure}

\begin{minipage}[b]{0.48\textwidth}
\includegraphics[width=0.8\textwidth]{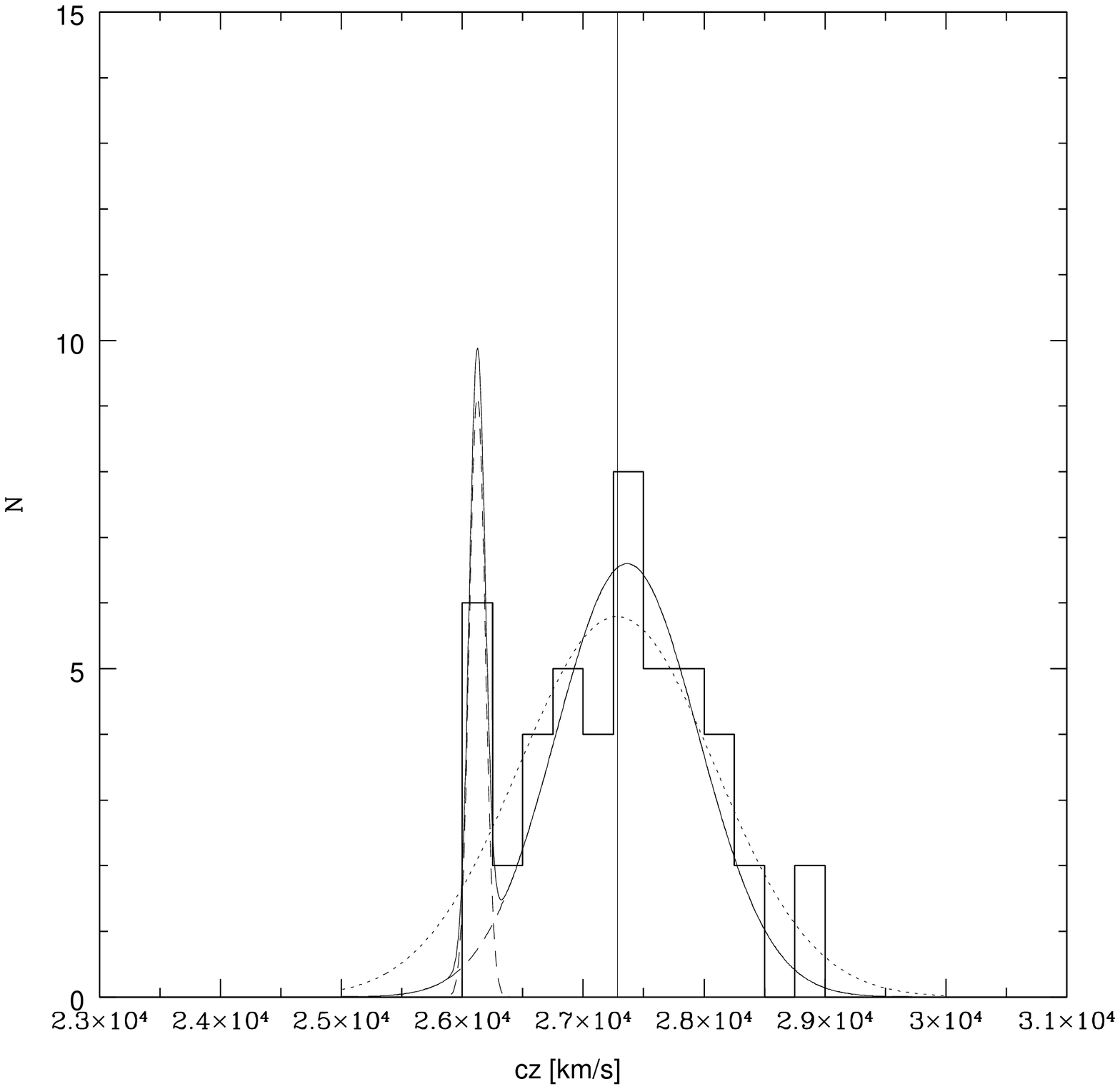}
 \caption{Velocity histogram of A2933 (binning of 250 km/s).
% obtained at ESO with the 3.6m
% telescope and EFOSC2 spectrograph with a binning of 250 km/s.
The best Gaussian fit for the whole distribution (dotted line) is centered on
the  vertical solid line which gives the location value. 
Location and scale
of the Gaussian (i.e.mean velocity and velocity dispersion) were estimated
with ROSTAT. 
We also show the two Gaussian functions corresponding to partition 1 and 2 of the best mixture 3 partitions fit by EMMIX 
(dashed lines) and the composite function (solid line). 
%The gaussian corresponding to partition 3, corresponding to the two high veloci%ty objects is not plotted.
}
\label{fig:A2933_histo_emmix}
%   \vspace{1cm}
   \end{minipage}

\begin{minipage}[b]{0.48\textwidth}
   \hspace{-0.9cm}
\includegraphics[width=0.95\textwidth]{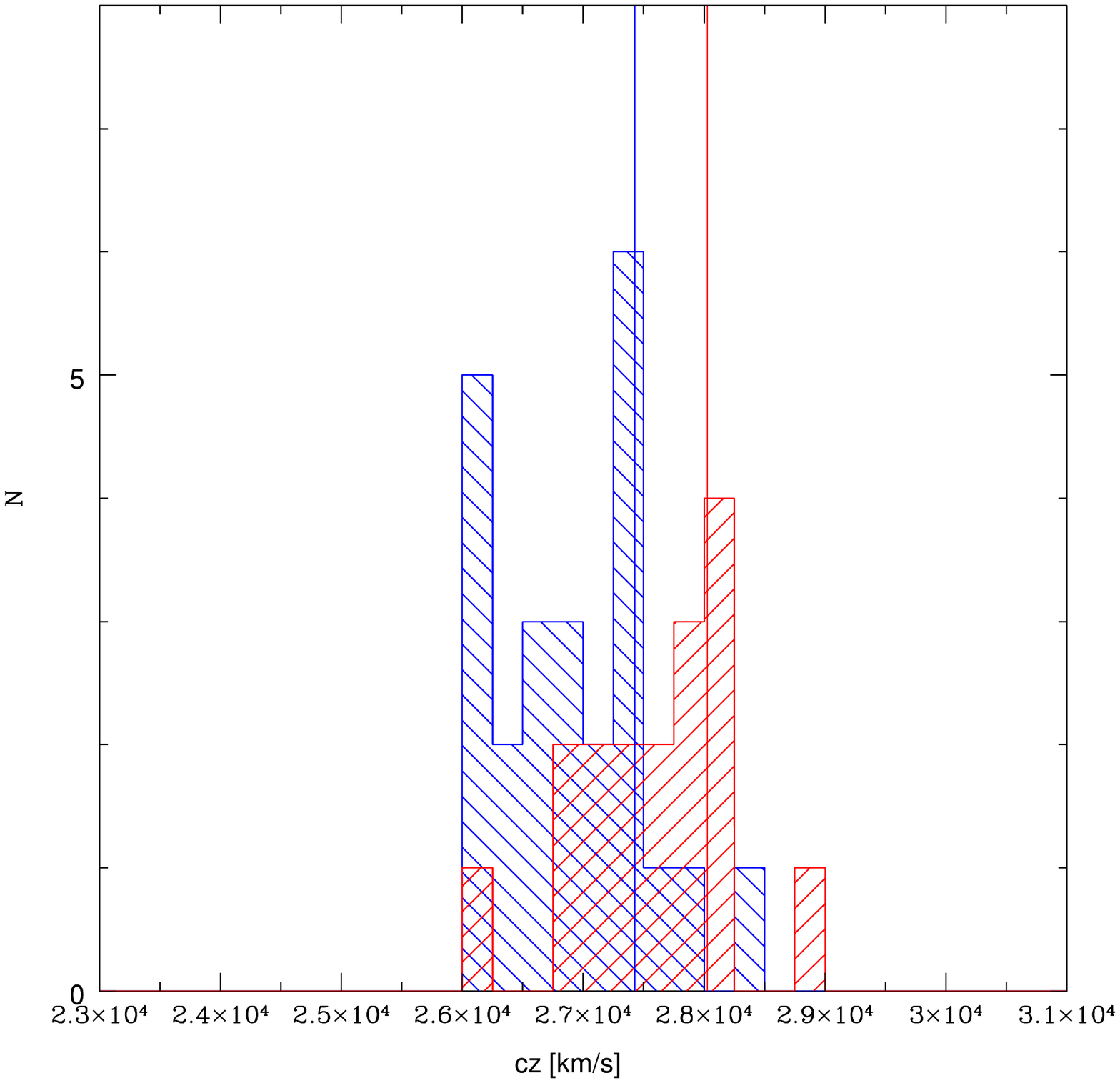}
\caption{Velocity histograms of A2933N (blue) and A2933S (red).
For each subcluster we included galaxies within a radius of 3.5' from its  
X-ray center. Solid lines: velocities of the brightest
 galaxy in the Northern (blue) and Southern (red) subclusters.}
\label{fig:A2933_histo_subclus}
\end{minipage}

\end{figure}

The velocity histogram (plotted in Fig \ref{fig:A2933_histo_emmix}) clearly 
shows a multi--modal distribution, with a sharp peak at $\sim 26\,200$ km/s, 
and a more dispersed structure, with two wider peaks at 
$\sim 26\,800$ km/s and $\sim 27\,500$ km/s.
In Fig.\ref{A2933_spectro}, cluster members are indicated  by
different symbols and colors for galaxies with velocities 
within each of the three peaks in the histogram. 
Most of the galaxies with velocities located in the sharp peak at 
$\sim 26\,200$ km/s (purple squares) are located in A2933N. Moreover, 
 the region of A2933S, including its southern extension, is 
populated mostly by galaxies in the highest velocity peak at $\sim 27\,500$ km/s, 
while objects with velocities within the peak at $\sim 26\,800$ km/s are
mainly in the A2933N region. 

To quantitatively assess the significance of the multi-modality
in the velocity distribution suggested by the histogram, 
we applied our set of statistical tests
(running a set of tests is useful because they have different sensitivities
to subclustering; see  Pinkney et al. 1996).
Ten out of 10 of the normality tests provided by ROSTAT do not find
significant deviations from Gaussianity (see 
Table \ref{table::rostat}). However, the Dip-test excludes unimodality at 
more than $2 \%$ significance. 

We finally applied EMMIX, where the data were fitted without a
priori constraints on the covariance matrix. 
A three Gaussian mixture inferred a P-value of 0.03, indicating that 
the Gaussian hypothesis is excluded at a very high significance level. 
This mixture fits  the main structure with a Gaussian of  mean 
$V = 27\,364~$km/s and dispersion $\sigma =501~$km/s (partition 1), the sharp 
peak at lower velocity with a Gaussian having mean $V \sim 26\,145~$km/s 
and a very low dispersion $\sigma = 65~$km/s (partition 2), and the small
excess at  high velocities (partition 3). 
Results are listed in Table \ref{table::emmix}. 
The sharp low velocity excess consists of six galaxies:
all but one are located in the northern subcluster 
(purple squares in Fig. \ref{A2933_spectro}). These galaxies are likely 
to form a low velocity group probably infalling into A2933N. 
In the following, we  exclude partition 3 (with only 2 objects) 
from the analysis. 
No significant results are obtained with EMMIX when fitting with mixture with a 
higher partitioning  to the data.

%
% FIGURE 3 & FIGURE 4
% 2 VELOCITY HISTOGRAMS OF A2933

\subsection{X-ray/optical combined analysis}

The projected density maps of A2933 for galaxies and for gas provide  similar 
views of the cluster (Fig.\ref{fig:OLT_a2933}): 
the X-ray and optical isocontours of the NW and SE subclusters, A2933N and 
A2933S, closely correspond and are centered on the 
cold cores in the temperature map (Fig.\ref{fig:OLT_a2933}). 
The X-ray contours are roughly centered on the BCGs for both the NW and SE 
subclusters. These results suggest that the system is in 
a pre-merger phase. 
However, in the case of the NW subcluster, the BCG seems slightly offset 
eastwards with respect to the X-ray centroid. The inner X-ray isocontours 
and the temperature 
map are elongated along a SE-NW axis connecting the two brightest galaxies 
of A2933N. 

As the density distribution of galaxies shows a clear bimodal behavior, we 
analyzed separately the velocity distributions 
of A2933N and A2933S, selecting the galaxies within a radius of 3.5' from 
each subcluster center (see Fig.\ref{A2933_spectro}). 
The corresponding histograms are displayed in 
Fig.\ref{fig:A2933_histo_subclus}. 
% with different color encoding. 
The velocity distributions of the two subclusters are offset from each other, 
the NW 
subcluster  lying at lower velocities than the SE 
subcluster, but there is significant overlap. Moreover, they 
are quite different. 
A2933N has a multi-peak structure with the sharp excess at low velocities 
previously identified with EMMIX, while A2933S 
has a more continuous distribution with an extended low velocity tail 
that falls in the velocity range of A2933N. 

\begin{figure}
\begin{minipage}[b]{0.48\textwidth}
\hspace{-0.9cm}
\includegraphics[width=0.95\textwidth]{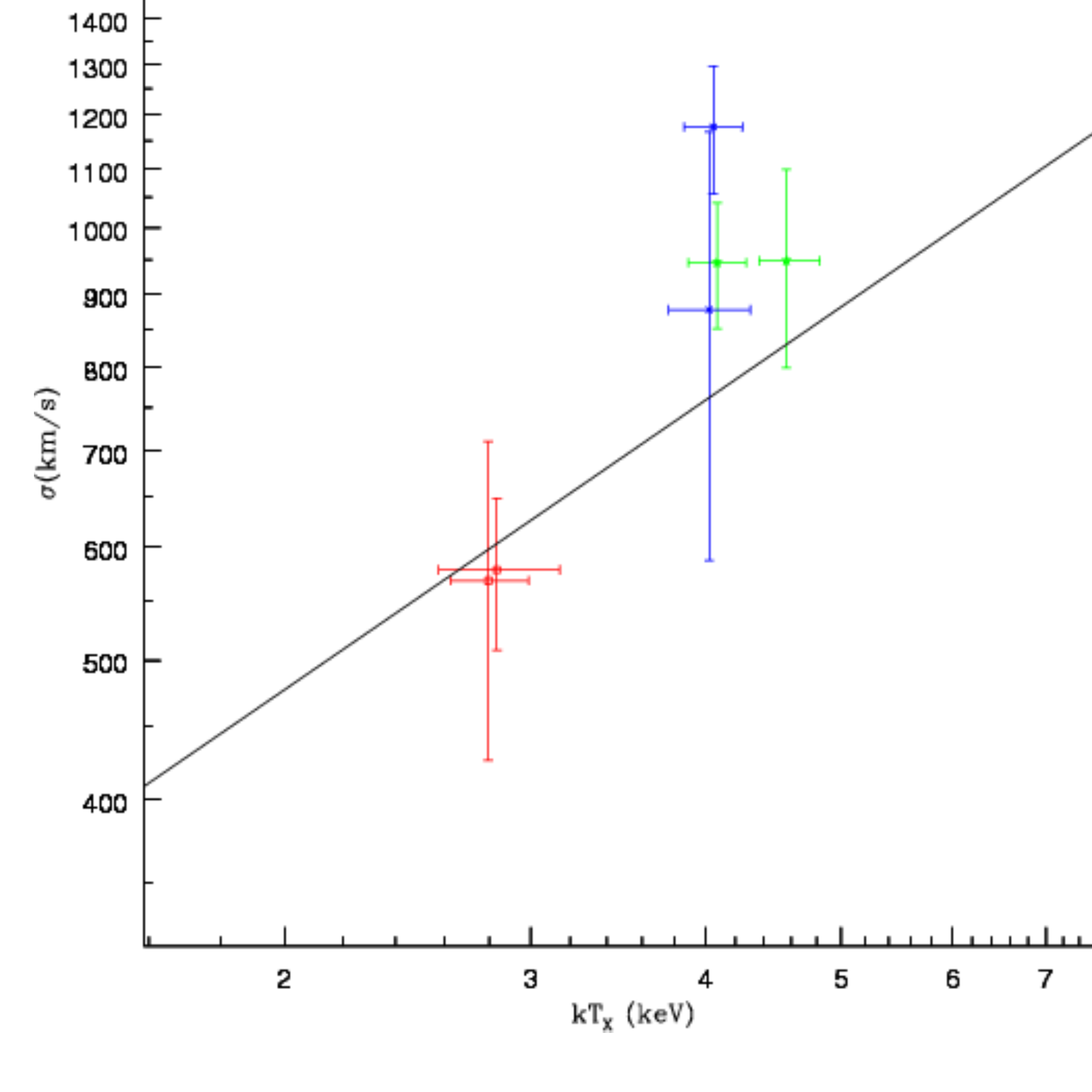}
\caption{The $\sigma-T_{X}$ relation for the subclusters A2933N and A2933S (red), 
A2440B and A2440C (green), A2384N and A2384S(blue). The straight line is the 
relation  $\sigma = 10^{2.47\pm 0.08}~T ^{0.67\pm 0.09}$ (Wu et al. 1998).}
%One can note that sub-clusters in the pre-merger A2933 lie on the general relat%ion, while in the case of post-mergers A2440 and A2384, the velocity dispersion% of the subclusters is high as compared to what expected from kT}
\label{fig::sigma_kt}
\end{minipage}
\end{figure}

Departure from normality was tested
when analyzing both subclusters  individually, and neither the normality tests nor 
the Dip test allowed us to significantly  exclude the 
hypothesis of a Gaussian velocity distribution.
The ROSTAT analysis of the two velocity distributions shows that the scales 
(velocity dispersions) 
are comparable for both subclusters.
These scales are in good agreement with those expected from the measured 
X-ray temperatures, assuming the typical scaling relation between $\sigma$ and kT 
(Lubin and Bahcall 1993, Girardi et al. 1998, Wu et al. 1998, 
see Fig.\ref{fig::sigma_kt}), suggesting that gas and galaxies are in equilibrium.

The dynamical analysis also confirms a significant 
offset for locations (mean velocities) of $\sim 730 \pm 210 $ km/s: A2933S 
has a significantly higher mean velocity than
A2933N ($\sim 27\,648$ km/s vs. $\sim 26\,919$ km/s). 
This trend is also followed by the velocities of the brightest cluster 
members: 
the BCG in A2933N (BCG1) and its close companion have comparable velocities 
($27\,350$ km/s and $27\,424$ km/s); the second brightest galaxy has a lower 
velocity ($26\,951$ km/s), 
while the brightest galaxy in A2933S (BCG2) has a higher velocity ($28\,028$ km/s). 
This implies that there is a significant offset of $\sim 600 \pm 140$ km/s 
between BCG1 and BCG2. 

The velocities of the two BCGs also differ from
the mean velocity of their host subcluster by $505\pm165$ km/s and 
$380\pm 200$ km/s for A2933N and A2933S, respectively, which is 
statistically significant for A2933N.

However, they coincide with the 
locations of the highest peaks in the velocity histograms of their respective 
subclusters A2933N and A2933S (excluding the sharp peak at $\sim 26\,200$ km/s),
and  these peaks are slightly shifted from the {\it mean} velocity of each subcluster, due to the skewness of 
the velocity distributions (Fig.\ref{fig:A2933_histo_subclus}). 
All these results suggest that the velocity 
distributions have started to mix with each other, but that the structure of the 
two subclusters is still not strongly affected; this implies
that the clusters have not yet crossed each other.

Therefore the combined X-ray--optical analysis indicates that 
A2933N and A2933S are in an advanced pre-merger stage. 
The presence of two very bright galaxies along an NE-SW axis, the elongations of 
the inner X-ray contours and  the temperature map along the E-W 
direction of A2933N, and the existence of a distinct low velocity, 
low dispersion velocity component, suggest that A2933N has also undergone 
some previous merging and is still accreting a small group.  
%The complex velocity and temperature structure of A2933N 
%suggests a dynamically young structure created by a recent merger.

\subsection{Dynamical analysis}

We now provide mass estimates for the subclusters and the total
mass of A2933 (analogous sections are devoted to A2440 and A2384).
For each system, we discuss the uncertainties in the 
mass estimates. An obvious uncertainty is related to the assumption of
dynamical or hydrostatic equilibrium. For example, velocity dispersions
might not simply reflect the subcluster potential and might be overestimated.
Another uncertainty is due to the limited field in which it is possible
to estimate the harmonic radius: the circle within which we estimate
this radius cannot intersect the nearby subcluster(s). As a consequence,
the virial radius might be underestimated. The mass may also be
underestimated  due to the pressure term.

\subsubsection{Mass estimates}\label{sec::mass_a2933}

From the previous analysis, A2933 is very likely in a pre-merger stage, 
in which subclusters begin to interact but cores have not yet crossed each 
other. We may then assume that the states of the two subclusters do not strongly
deviate from dynamical equilibrium, so that 
the virial theorem and the hydrostatic 
equilibrium can be safely applied to determine optical and X-ray mass 
estimates. 
These estimates will be used in the following  to 
constrain the parameters of the collision by means of a two-body analysis. 

In the optical, we estimated the harmonic radius of the two
subclusters from the projected
distribution of all galaxies with $R < 20$ and within a radius of 3.5 arcmin
from the BCG; from the harmonic radius we derived the virial radius and,
 in combination with the velocity dispersion, we estimated 
the virial mass. Finally, extrapolating the density profile, 
we estimated the mass $M_{200}$ corresponding to a density contrast 
$\bar{\rho}/\rho_c = 200$ 
(for more details see e.g. Maurogordato et al. 2008).
We find for A2933N $M_{200} = 2.0 \pm 0.4 \times 10^{14}$ $M_\odot$ and for
A2933S $M_{200} = (2.2 \pm 0.4) \times 10^{14}$ $M_\odot$, with a ratio
$M_{A2933S} / M_{A2933N} = 1.1 \pm 0.6$.

We can also estimate the mass ratio using the X--ray scaling relations. 
The mean X-ray temperatures of A2933N and A2933S are respectively
 $kT_{A2933N}=2.11\pm0.25 $keV  and $kT_{A2933S}=2.7\pm0.2$keV. 
Assuming $M\propto T_X^{3/2}$, we obtain a mass ratio between the two 
components around 1.5:1. 

On the other hand, when we estimate masses 
using the brightness and temperature profiles of each unit, assuming hydrostatic 
equilibrium for each subcluster, 
we find that $M_{A2933N}=2.6 \pm 0.4 \times 10^{14}M_\odot$ and 
$M_{A2933S}=2.76 \pm 1.5 \times 10^{14}M_\odot$ leading to a mass
ratio of $\frac{M_{A2933S}}{M_{A2933N}} = 1.2 \pm 0.6$. 

While the exact values could be systematically affected 
by deviations from equilibrium, there is a good consistency between the
optical and X--ray mass estimates, and we can conclude that the two 
subclusters have comparable masses, with a total mass for A2933 of  
$M_{200} \sim 5 \times 10^{14} M_\odot$. 

\begin{figure}
\centering %Fichier absent
\includegraphics[angle=0,width=0.9\columnwidth,keepaspectratio] 
{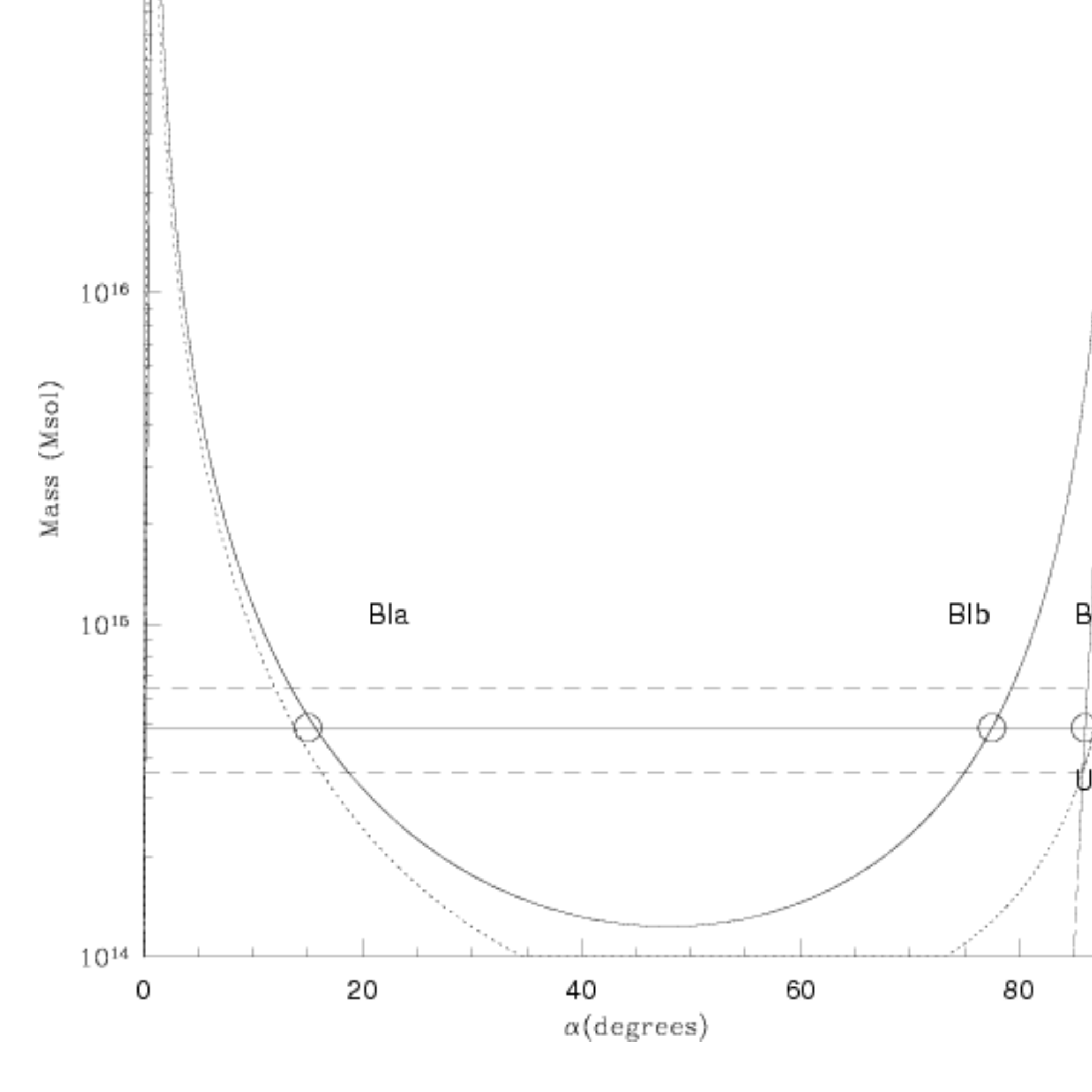}
\caption[]
{The sum of the virial masses of A2933N and A2933S
as a function of the projection angle $\alpha$. 
The horizontal lines show the mass estimate 
(full line) and errors (dashed lines) (see 
Sect.~\ref{sec::mass_a2933}). The projected distance of 0.75  Mpc and radial 
velocity difference at rest of 550 km/s are derived from the two BCG's 
location and velocities. The two systems were at zero separation 12.2 Gyr ago.
 The dotted line represents the Newtonian criterion for gravitational binding.
 Two bound incoming (BIa and BIb) and one bound outgoing (BO) 
are found compatible with our mass estimates.}
\label{fig:A2933_2body}
\end{figure}

\subsubsection{Two-body analysis}

We applied the two-body dynamical formalism (Beers et al. 1982; 
Gregory and Thomson 1984) to the two subclusters in A2933. One of the
variants of this method establishes a relationship
between the total mass $M$ of the bimodal cluster and the angle
$\alpha$ between the plane of the sky and the collision axis of the
two clumps (Barrena et al. 2002, Ferrari et al. 2005). The input parameters 
derived from the observations are the relative radial velocity between the two 
subclusters $V_r$ and their projected spatial separation $R_p$. 
For these observables, we chose to take the values 
of the projected spatial separation and the radial relative velocity (at rest)
 between the two BCGs ($R_p=0.75~$ Mpc and $V_r=550~$km/s), which are more clearly
defined than that of the subclusters. We obtained similar results when using 
values from the subcluster centroids. 
In addition, the formalism requires us to
define the time $t_0$ that has elapsed since the epoch of the last encounter
between the two subclusters. We tested different scenarios, including a
pre-merging scenario ($t_0 \sim 12.2$ Gyr, i.e. the age of the
Universe at the redshift of A2933, implying that the two systems have
never yet crossed each other), a recent ($t_0=0.5$ Gyr) and an older
($t_0=1.0$ Gyr) post-merger event. Considering the total mass of the
cluster derived from optical observations ($5 \times 10^{14} M_{\sun}$), 
the post-merger cases provide only one possible bound
outgoing solution (i.e. the two subclusters are moving 
apart). In the pre-merger case, Fig.\ref{fig:A2933_2body} reveals two possible
 bound incoming solutions, the first with $\alpha \sim  15^{\circ}$ (BIa)
and the second with  $\alpha \sim 75^{\circ}$ (BIb), 
and a bound outgoing solution (BO).
For BIa, the ``real'' separation between the two subclusters appears to be small 
($\sim 0.8$ Mpc) and the relative velocity high ($V \sim 2000$ km/s), 
while for BIb the two components appears to be more distant ($\sim 3.5$ Mpc) and 
the relative velocity comparable to the projected observed one 
(V$ \sim 560$ km/s). The results are listed in Table \ref{table::2body}.
% In the following we discuss the viability of these different solutions 
% taking into account the results of the X-ray/optical analysis.

\subsection{The scenario for A2933} 

The combined X-ray optical analysis has shown: 
a) two subclusters, visible in both X-ray and optical maps, and 
associated with cool cores detected in the temperature maps;
b) a hot region in between the two subclusters;
c) a perturbed velocity distribution, showing a  velocity offset (at rest) of $\sim 700$ km/s 
between the two subclusters.
This suggests that the interaction between the two subclusters is in its
initial phase.
This lead us to exclude the post-merger cases in the two-body analysis, and
to examine the pre-merger one. Solutions with large values of $\alpha$,
 BIb, and UO are quite improbable since they would imply large 
spatial separations
between the subclusters ($\sim $ 3.5 Mpc and $11$ Mpc, respectively), which is 
unlikely taken into
account the clear signs of interaction between the two merging units.
The most likely solution for A2933 is thus (BIa), i.e. a two-body pre-merger 
nearly on the plane of the sky ($\alpha \sim 15^{\circ}$). 
The high relative velocity  ($\sim 2000$ km/s) between the two components 
implied by this solution is a common feature in merging clusters.

\begin{table*}
\begin{center}  
\small
\begin{tabular}{cccccccccccc}
\hline
\hline
Name &RA(BCG)  &DEC(BCG)  &$\theta_p$  &$C_{BI}$(km/s)   &$S_{BI}$(km/s)  &Nobj &$V_{BCG}$(km/s) &Ntest  &Dip Test &$P_{value}$ &$T_X(KeV)$\\
\hline
A2933 &--- &--- &15 &27\,281$\pm$ 103 &682  $\pm$ 65 &47  &--- &0  &0.0340  &0.02 &---\\
A2933N  &01 40 35.520  &-54 30 54.00  &3.5 &26\,919$\pm$ 131  &578$\pm$ 70 &24 &27\,424$\pm$ 100 &0 &0.056 &0.12 &$2.83_{-0.26}^{0.31}$\\
A2933S  &01 40 59.520  &-54 37 26.40  &3.5 &27\,648$\pm$ 166 &568$\pm$ 142 &17 &28\,028$\pm$ 100 &0 &0.071 &0.15 &$2.80_{ -0.17}^{0.19}$ \\
\hline
A2440 &---  &--- &15 & 27\,251 $\pm$ 93 & 940$\pm$70 &103 &--- &0 &0.022 &0.01 &---\\
A2440A &22 24 13.663  &-01 31 34.34 &2.5 & 28\,005 $\pm$ 75 & 178$\pm$90 &10 &27\,925$\pm$ 100 &10 &0.067 &0.03 &$---$ \\
A2440B &22 23 56.941   &-01 34 59.78 &2.5 & 27\,212$\pm$ 195 & 946$\pm$95 &29 &27\,032$\pm$16 &1 &0.050 &0.1 &$4.07_{-0.19}^{0.20}$ \\
A2440C &22 23 47.814  &-01 39 01.14  &2.5 & 27\,234 $\pm$ 188 & 949$\pm$150 &25 &26\,816$\pm$19 &0 &0.059 &0.15 &$4.57_{-0.20}^{0.26}$ \\
\hline
A2384 &--- &--- &15 &28\,263$\pm$ 154 &1114$\pm$ 106 &56 &--- &0   &0.038   &0.1
&---\\
A2384N &21 52 21.962 &-19 32 48.65  &5 &28\,139$\pm$ 182 &1176$\pm$ 120 &44 &27\,602$\pm$55 &0  &0.043  &0.1 &$4.05_{-0.19}^{0.20}$\\
A2384S &21 52 09.556  &-19 43 23.71 &5  &28\,814$\pm$ 241 &877$\pm$ 290 &11 &28\,696$\pm$53 &0  &0.066  &0.03 &$4.02_{-0.26}^{0.28}$\\
\hline
\hline
\end{tabular}
\caption{Characteristics of the clusters and subclusters. Location, scale, and X-ray mean temperatures of the different subclusters are computed in circles of $\theta_p$ arcmin centered on the position of the BCG (associated with the corresponding  X-ray maximum). The measured velocity of the BCG is also listed. }
\label{table::rostat}
\end{center}  
\end{table*}

\begin{table*}
\begin{center}  
\small
\begin{tabular}{cccccc}
\hline
\hline
Name &part &v1 (km/s)& $S_{BI}$ (km/s)&$N_g$ &Allocation rate \\
\hline
A2933 &1 &$27\,364 \pm 84$ &$501 \pm 45$  &39 &0.984 \\
A2933 &2 &$26\,145 \pm 27$ &$65 \pm 25$   &6  &0.999 \\
A2933 &3 &$28\,873 \pm -$  &-  &2  &1.0 \\
\hline
A2440 &1 &$27\,080 \pm 96$ &$889 \pm 62$ &96 &0.95 \\
A2440 &2 &$28\,058 \pm 25$ &$33 \pm 25$ &5  &0.30 \\
A2440 &3 &$30\,360 \pm -$ &- &2  &1.0 \\
\hline
A2384 &1 &$29\,760 \pm 122$ &$365 \pm 75$ &13 &0.968 \\
A2384 &2 &$28\,546 \pm 55$  &$200 \pm 27$  &20 &0.974 \\
A2384 &3 &$27\,312 \pm 168$ &$654 \pm 135$ &23 &0.833 \\
\hline
\hline
\end{tabular}
\caption{Results of the mixture of Gaussian with EMMIX. Errors have been estimated with ROSTAT.}
\label{table::emmix}
\end{center}  
\end{table*}

\begin{table*}
\begin{center}
\small
\begin{tabular}{ccccccc}
\hline
\hline
cluster  &$t_0$        &solution  &$\alpha$(deg) & R(Mpc)  & Rm(Mpc)  &V(km/s)\\
\hline
\hline
A2933N/S &12.2 Gyr  &BIa       &15       &0.8     &4.1       &2120\\
          &         &BIb       &78        &3.5     &4.7       &560\\
          &         &BO        &86        &10.9     &47.      &551\\    
\hline
A2933N/S &0.5Gyr    &BO        &47       &1.1    &1.5      &1055\\ 
\hline
A2933N/S &1.0Gyr    &BO        &65       &1.8    &2.5      &862\\
\hline
\hline
A2440A/B+C &12.2Gyr &BIa       &11.37     &0.61   &5.3      &3700\\
\hline
           &        &BIb       &82.4     &4.5    &6.1      &740\\
\hline
           &        &BO        &87.5     &14.3   &75       &730\\    
\hline
\hline
A2384N/S &12.2Gyr    &BIa        &17       &1.2     &6.5       &3440\\         
         &           &BIb        &77       &5.0    &7.3      &1030\\         
\hline
A2384N/S &0.5Gyr    &BO        &46       &1.6     &2.0       &1390\\         
\hline
A2384N/S &1.0Gyr    &BO        &65       &2.6     &3.2       &1110\\   
\hline
A2384N/S &2.0Gyr    &BO        &75       &4.4     &6.0       &1030\\  
         &          &BIa       &25       &1.2     &2.2       &2354\\  
         &          &BIb       &53       &1.9     &2.3       &1240\\        
\hline
\hline

\end{tabular}
\caption[]
{Two--body model solutions for the A2933N-A2933S, A2440A-A2440(B+C),
 and A2384N-A2384S systems. Various possibilities are considered for
the evolutionary phase: the pre-merger case 
($t_0 = 12.2 Gyr$), and post-mergers seen at $t_0$ after the
first passage. For each solution, we derive the
angle $\alpha$ between the line connecting the two components and the plane of
 the sky, the spatial separation of the subclusters $R$, their separation 
at maximum expansion $R_m$ and their relative velocity $V$.}
\label{table::2body}
\end{center}
\end{table*}

%==============================================================================
\section{A2440}

%==============================================================================

Mohr et al. (1996) performed a combined X-ray and optical analysis of A2440, 
based on optical imaging  and spectroscopy (48 redshifts), and Einstein 
X-ray data. Our work significantly extends this study, being based on a 
larger spectroscopic sample (more than double the total number of 
redshifts) and new X-ray data, including the gas temperature maps.

\subsection{X-ray gas morphology and thermal structures}

\begin{figure*}[ht]
\centering
\hspace{-0.5cm}
\includegraphics[width=\textwidth]{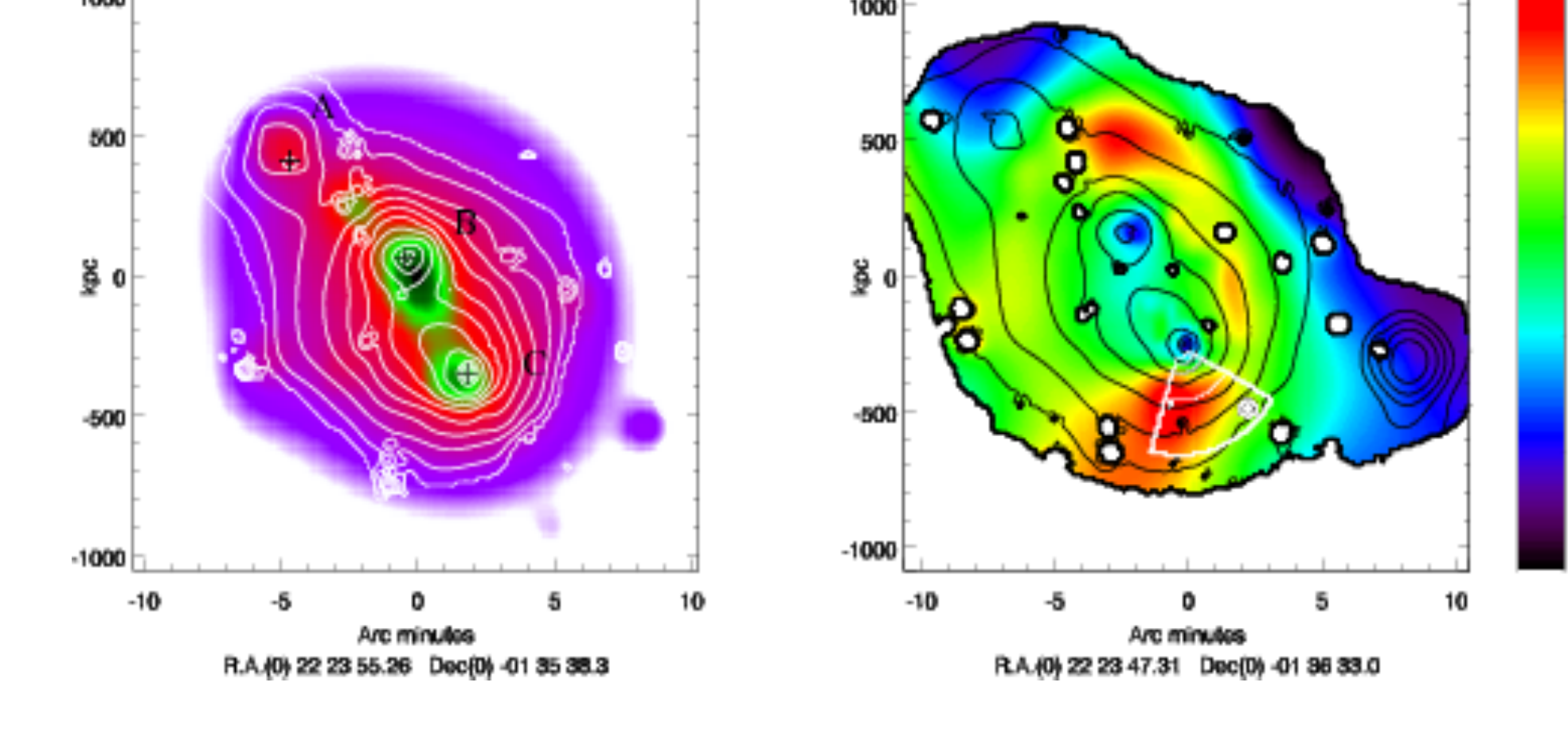}
\caption{A2440: Left: Galaxy density map 
(mag. cuts: R$<$19) overlaid on the X-ray luminosity contours (EPIC-XMM 
 counts in the .5-2.5 keV band corrected for background and vignetting). Black 
 crosses indicate the BCG positions. Right: ICM temperature map overlaid 
 on the X-ray luminosity contours 
 (EPIC-XMM data analyzed through wavelet spectral-imaging, 
see Bourdin et al. 2004, 2008 for details).}
\label{fig:OLT_a2440}
\end{figure*}

\begin{figure*}[ht]
\centering
\hspace{-0.5cm}
\includegraphics[width=\textwidth]{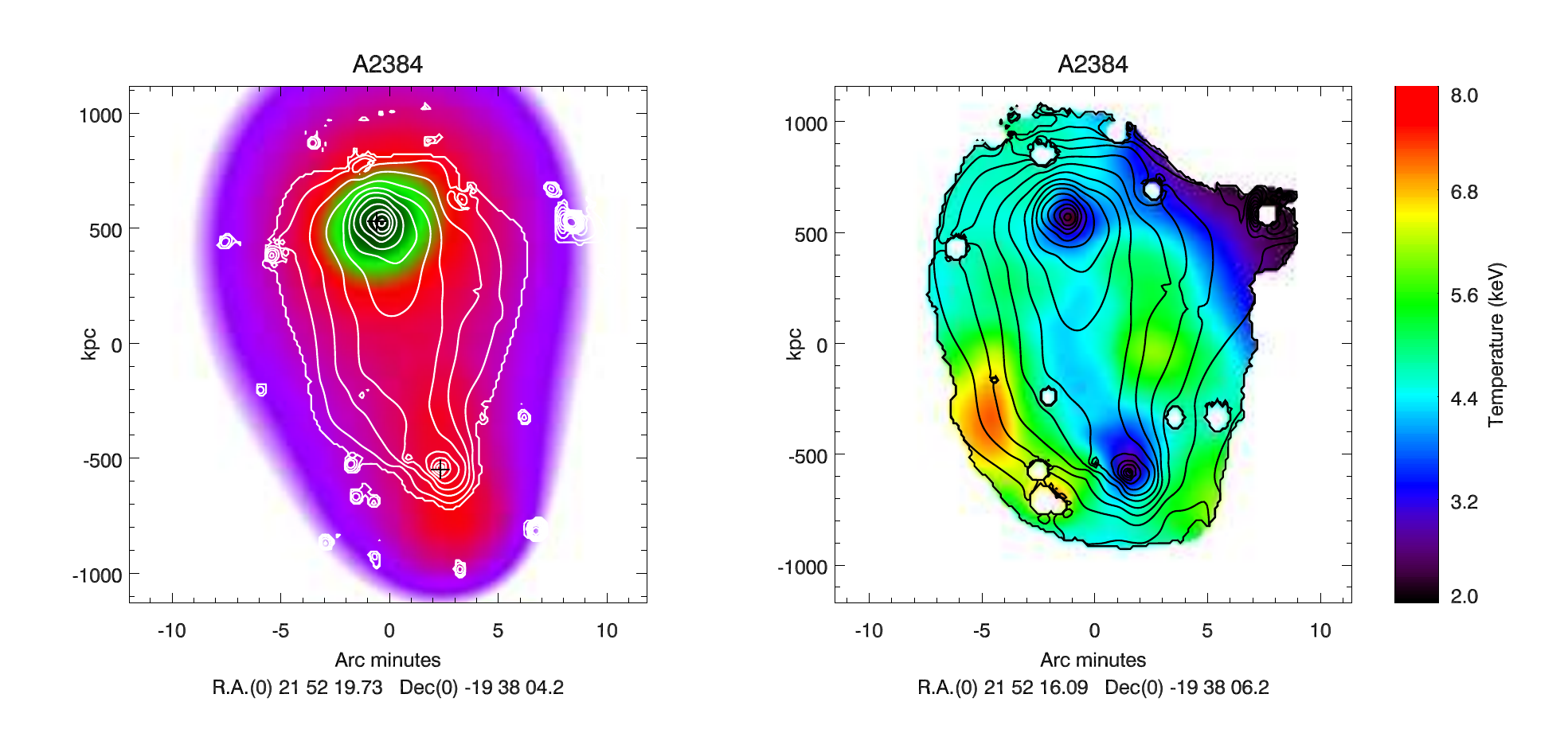}
\caption{Same as Fig. \ref{fig:OLT_a2440} for A2384.}
\label{fig:OLT_a2384}
\end{figure*}
The X-ray observations detect a gas emission with an elongated structure 
connecting two brightness peaks (B) and (C) to a northern and a less 
luminous component (A) (Fig. \ref{fig:OLT_a2440}). 
The two brightness peaks are associated with cool cores 
(kT $\simeq$ 2.8-3.5 keV), and are surrounded by hotter gas 
(kT $\simeq$ 5 keV); a significant estimate of the temperature 
of the (A) component is not possible due to the strong contamination 
from the innermost cluster regions. 

The southern emission peak (C) is delimited by gas brightness and temperature 
jumps across the white sector shown in Fig.\ref{fig:OLT_a2440}.  
As discussed above, these discontinuities are signatures of a cold front 
delimiting the edge of the southern cold core.

\begin{figure*}
\centering
\includegraphics[angle=0,width=\textwidth]{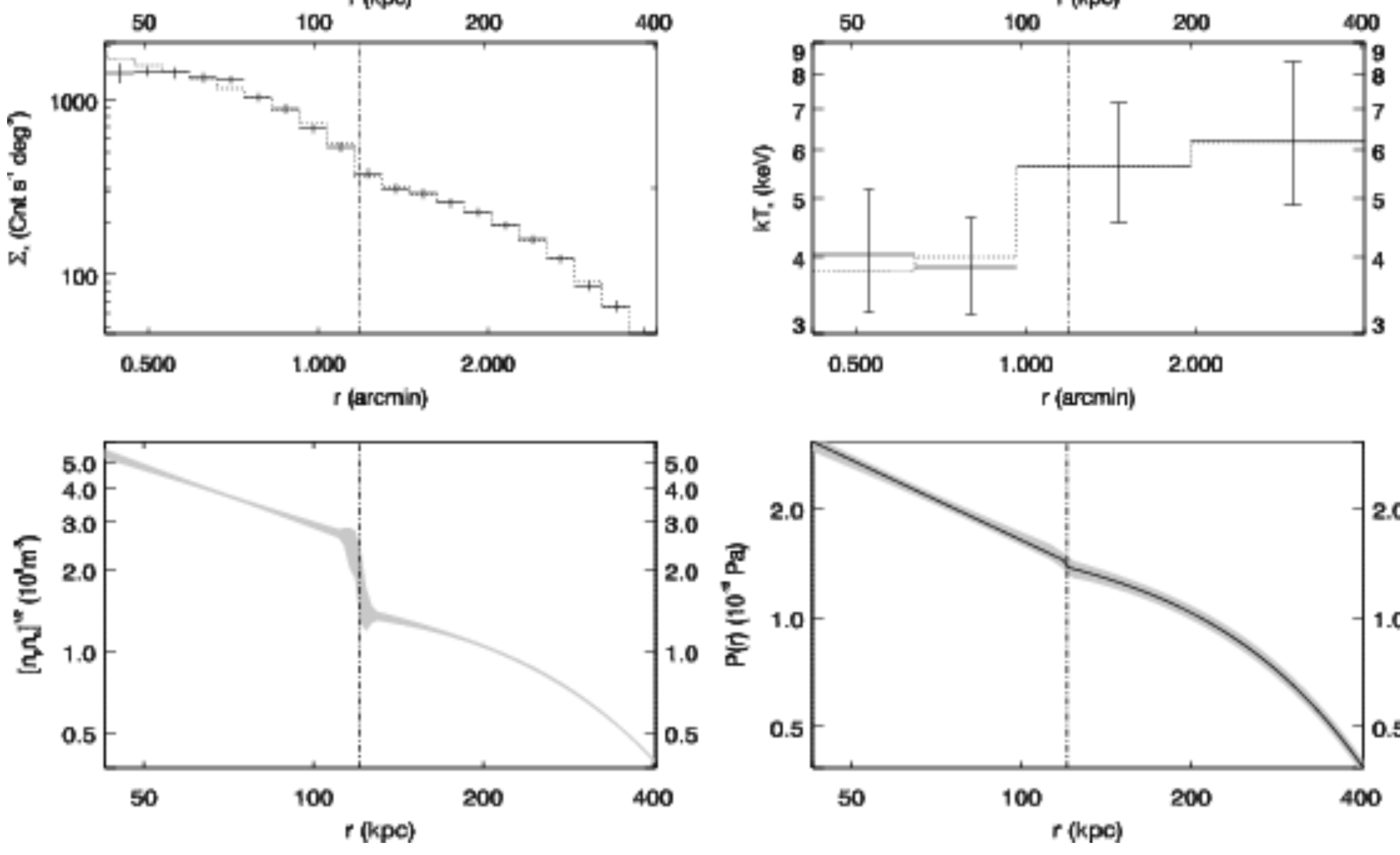}

\caption{Top: Gas brightness and projected temperature profiles observable 
across the white sector in Fig. \ref{fig:OLT_a2440} for A2440. 
Bottom: ICM density and pressure profile estimated above the Southern gas 
clump in A2440.
These profiles reveal a density discontinuity at constant pressure or 
"cold front" feature (see vertical dashed lines on the plots).}
\label{fig:cf_a2440}
%\vspace{-1.3cm}
\end{figure*}

To investigate the nature of the gas density jump observable across
the white sector in Fig.\ref{fig:OLT_a2440}, we model the gas 3D structure
with disrupted density and temperature profiles
(see Eqs. (20) and (21) of Bourdin et al. 2008). As shown
in Fig.\ref{fig:cf_a2440}, fitting the slope, jump position, and amplitude
of these functions reveals a discontinuity in gas density, 
$D_{n}=1.82\pm0.06$, and 3D temperature,
$D_\mathrm{T}=1.76^{+0.08} _{-0.06} $, located at 110 kpc to the south of the 
cool core.
The pressure continuity that is measurable across the jump
$\left(D_\mathrm{P} = D_\mathrm{T} / D_{n} =  0.97 \pm 0.05 \right)$ 
identifies the discontinuity as a
cold front. As already observed in various interacting systems,
this cold front is likely to delimit a stripped cold core moving outwards
from the hotter embedding ICM.

%
% FIGURE 10
% OPTICAL IMAGE OF A2440
%
\begin{figure}
\centering
\includegraphics[angle=0,keepaspectratio,width=\columnwidth]{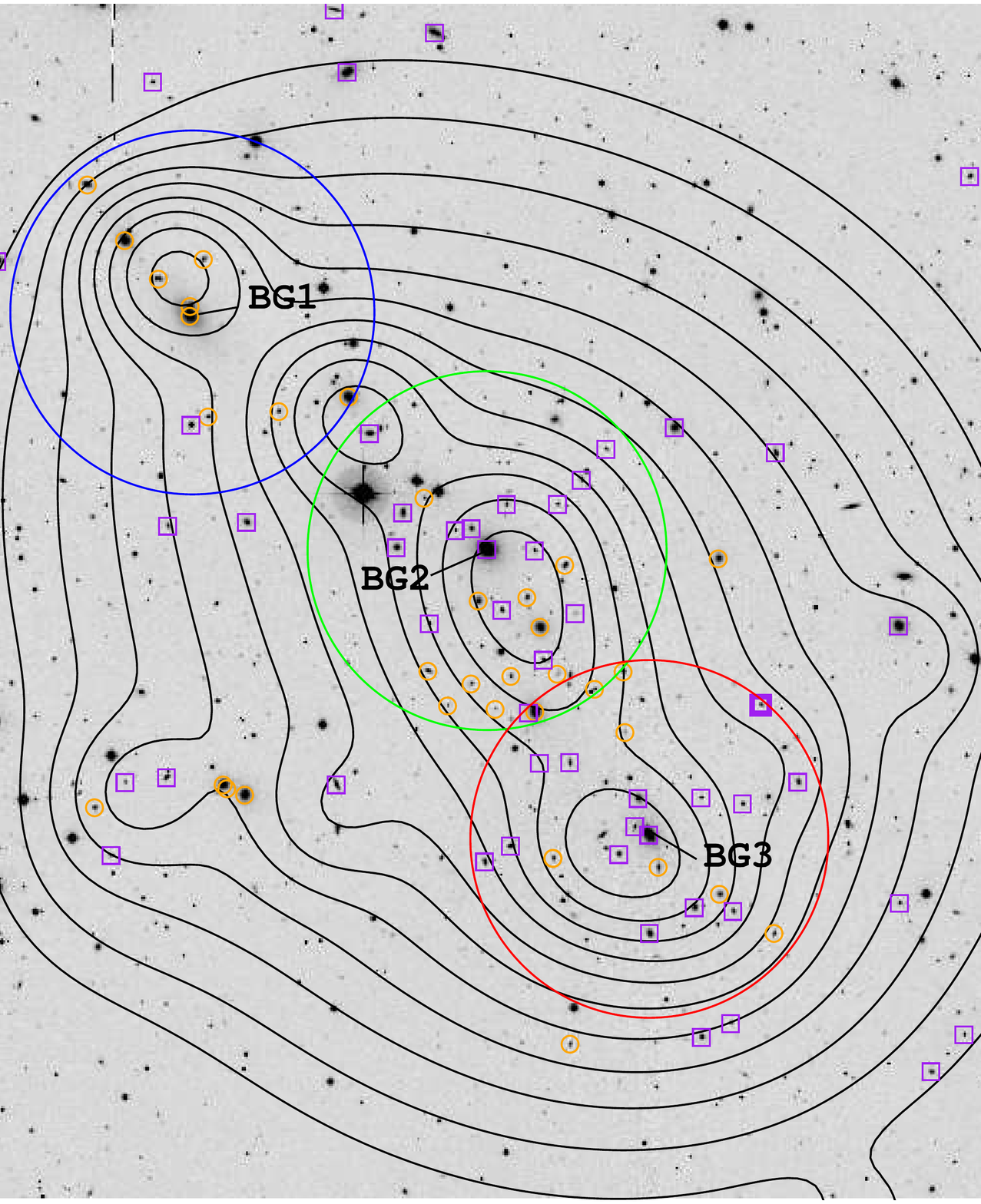}
\caption[]
{WFI R--band image of the A2440 field (12'x14'). 
Galaxies identified as cluster members from spectroscopy (Flag 0 and 1) are
marked with different symbols according to the velocity range 
(purple squares: [24000,27600] km/s; orange circles [27600,30000] km/s). 
The isocontours of 
red sequence galaxy density maps with magnitude limit $R<19$ 
are superimposed. The velocity histograms of the three regions 
with 2.5 arcmin radius delimited by circles 
(corresponding to the three subclusters A2440A, A2440B, 
and A2440C) are displayed in Fig.\ref{fig:A2440_histo_subclus}. 
North is up and East is to the left.}
\label{A2440_spectro}
\end{figure}

%
% FIGURE
% 2 VELOCITY HISTOGRAMS OF A2440
%
\begin{figure}

\begin{minipage}[b]{0.45\textwidth}
\includegraphics[angle=0,width=0.95\textwidth]{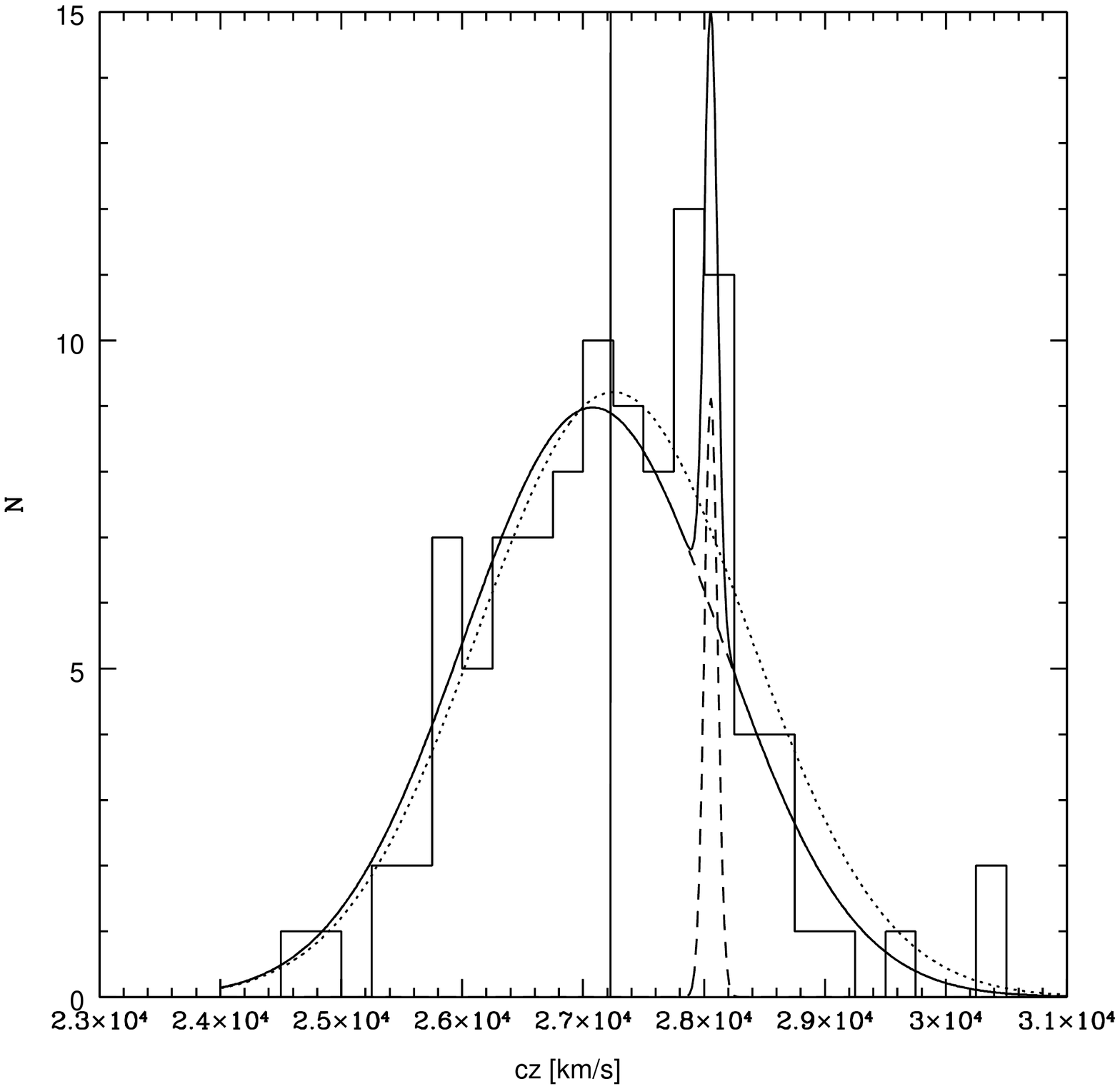}
\caption[]{Velocity histogram of A2440
% galaxy velocities obtained at ESO with the 3.6m
% telescope and EFOSC2 spectrograph, and collected from the literature with a 
(binning of 250 km/s). The best Gaussian fit for the whole distribution
(dotted line) is centered on the vertical solid line which gives the location
value. We also show the two Gaussian functions (dashed lines)
corresponding to partitions 1 and 2 in the best 3 partition 
mixture model by EMMIX and the composite function 
(solid line).}
\label{fig:A2440_histo_emmix}
   \end{minipage}

\hspace{0.7cm}

\begin{minipage}[b]{0.45\textwidth}
\includegraphics[angle=0,width=0.95\textwidth]{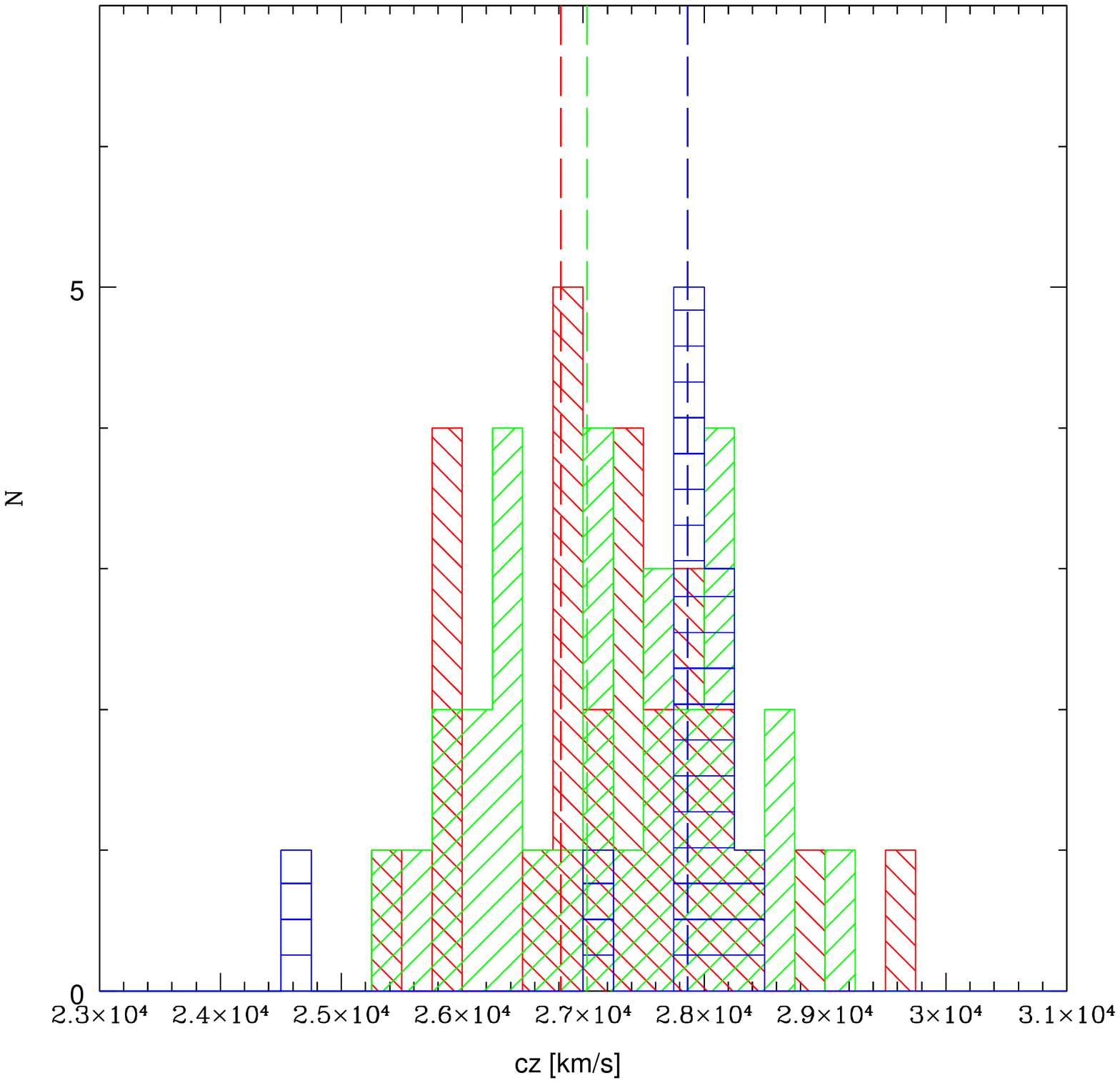}
\caption[]{Velocity histograms of A2440A (blue), A2440B (green) and
A2440C (red). For each subcluster we included galaxies within a radius of 
3' from its center.
Dotted lines: velocities of the brightest galaxies in A2440A (blue),
A2440B (green) and A2440C (red).}
\label{fig:A2440_histo_subclus}
\end{minipage}

\end{figure}

\subsection{Galaxy density distribution }

The galaxy density maps computed for RS galaxies for different magnitude cuts (Fig.\ref{isodens_mag}) 
show that the general structure of A2440 is strongly elongated along a NE/SW 
axis and includes multiple clumps. At bright magnitudes  ($R<18$),
four groups are detected: the two most significant subclusters B and C 
are identified as the main cluster, and two fainter ones are detected at 
the NE extent: A1 and A2 . When including faint galaxies, the two components 
corresponding to the main cluster B and C progressively merge to appear as a 
single dominant
elongated structure ($R < 21$), while the NE groups A1 and A2 are less visible
and a new (but faint) component A3 is detected on the western side. 

The three brightest galaxies are located near the three density peaks A1, B, 
and C identified at $R < 18$ and $R < 19$ 
(Fig. \ref{A2440_spectro}).

There is a strong alignment between the main axis of the bright galaxies, 
the position angle of the subclusters and the NE/SW axis of the whole cluster. 

\subsection{Galaxy velocity distribution}

The first analyses of the dynamics of A2440 were performed
by Beers et al. (1991) and Mohr et al. (1996).
We now have a much larger dataset:  
from our EFOSC2 observations, 97 redshifts have been obtained in the 30' by 30' field centered on A2440. We merged this catalog with redshifts available in the literature, which
led to the final redshift catalog of 150 objects (with 10 twice observed objects).
%%% ADD
A comparison with 10 measured redshifts available
in the literature gives a mean difference of 30 km/s and 
a standard deviation of 60 km/s.
%%% END ADD
Restricting this sample to cluster members, we are left with 103 objects. 
The velocity histogram is shown in Fig.\ref{fig:A2440_histo_emmix}.

In Fig.\ref{A2440_spectro}, cluster members are indicated by
purple squares and orange circles when belonging to one of the two major 
velocity peaks in the 
histogram at $\sim 27\,000$ km/s and 28\,000 km/s respectively.
One can see that galaxies belonging to both velocity peaks 
populate subclusters B and C, while all galaxies in subcluster A1 
belong to the highest velocity peak. 
%However, the velocity distribution could still have a unique peak with 
%an excess at $\approx 27800kms^{-1}$. 

Measurements of location and scale of the velocity distribution with the biweight estimator give 
$C_{BI} = 27\,251 \pm 93 kms^{-1}$ and $S_{BI} = 940 \pm 70 kms^{-1}$. 
 While none of the  10 ROSTAT normality tests identify 
significant deviations from Gaussianity, 
the Dip test rejects unimodality at the 1 percent level 
(Table \ref{table::rostat}) .

EMMIX fits a mixture with two partitions isolating the main component and an 
excess of two galaxies at $V \sim 28\,058$ km/s with a very significant 
P-value (0.01). With three partitions,  EMMIX isolates the main component 
with mean velocity $V \sim 27\,080$ km/s and  velocity dispersion 
$\sigma = 890$ km/s (partition 1),  a very
sharp Gaussian ($\sigma \sim 33$ km/s), corresponding to the previously
mentioned excess with mean velocity $V \sim 28\,058$ km/s (partition 2), 
and a group of two galaxies at $\sim 30\,337$ km/s (partition 3), which are not considered in the following. 
The best-fit Gaussian functions (partition 1 and 2) are 
plotted in Fig. \ref{fig:A2440_histo_emmix}. However, the relatively high value of 
$P_{value}$ obtained (0.23) indicates that this three partition fit is less 
significant than the two partition one. No significant results are obtained with 
EMMIX when fitting with a larger number of partitions.

\subsection {A2440: X-ray/optical combined analysis}

Examining Fig.\ref{fig:OLT_a2440}, where we show the X-ray isocontours and 
the galaxy projected density distribution, we see that the central X--ray 
bimodal structure (B+C) is nearly coincident with the optical bimodal 
structure (B+C) in the density map at $R<19$ (and coincident with that 
at $R<18$).
The X-ray A component is coincident with the NE subcluster A1 identified
in the galaxy density maps, while the subclusters A2 and A3 are not 
detected in X--rays. 
The three X-ray maxima A, B, and C are 
well centered on the three corresponding BCGs (Fig.\ref{fig:OLT_a2440}). 
We have seen that at fainter magnitudes the optical bimodal structure 
merges into one structure: its density peak is centered   
between the X-ray maxima B and C. These three subclusters, identified in X--rays
 and with optical counterparts, are referred to as A2440A, A2440B,
 and A2440C.

To analyze the velocity distribution in A2440A, A2440B, and A2440C, 
we  defined three subsamples including
galaxies within circular regions of radius 2.5 arcmin (the largest
compatible with negligible overlap)
and centered on the corresponding X--ray maxima, which coincide with the BCGs 
positions. These circular regions are shown in Fig.\ref{A2440_spectro}.

In Fig.\ref{fig:A2440_histo_subclus}, we show the corresponding
velocity histograms with different colors 
(blue, green, and red for A2440A, A2440B, and A2440C, respectively).
The histograms indicate that most galaxies in A are in the
velocity bin at $\sim 28\,000$ km/s (EMMIX partition 2). 
Within the limits of the small number
of counts per bin, there is no evidence of another significant segregation
in velocity for galaxies in the regions $B$ and $C$:
the velocity distribution seems spatially mixed. 

The results of the ROSTAT analysis for the regions corresponding to the three 
X-ray subclusters are listed in Table \ref{table::rostat}. Both 
the mean velocity and velocity dispersions of the $B$ and $C$ regions 
are comparable, and consistent with the global cluster values. 
This suggests two possible alternatives: either A2440B and A2440C have not yet 
interacted, and the merging is occurring 
in the plane of the sky, or they have already crossed, but are seen just after
core passage. The velocity of the BCG in the $B$ region, BG2, 
is close to the velocity of its host subcluster, 
while BG3 in the $C$ region has a ($\sim 2 \sigma)$ velocity offset 
of ($ 418\pm 188~kms^{-1}$).
The velocity dispersions of the two main components 
A2440B and A2440C are higher than the values expected from their 
X-ray temperature (see Fig.\ref{fig::sigma_kt}), but the deviation 
from the $\sigma-T_X$ relation is not statistically significant.

The mean velocity of A2440A is $\approx 750kms^{-1}$ higher than that of the 
main cluster  at a $3 \sigma$ level. The small velocity dispersion of $A$ is more 
typical of a group.
The velocity of the brightest galaxy in $A$, $VG1 = 27\,925 $ kms$^{-1}$, 
is consistent with the mean velocity of its subcluster, implying that it is at 
rest in the potential well.  This dynamical analysis is in very good agreement 
with that obtained by Beers et al. (1991) and Mohr et al. (1996). 

These results show that A2440A is a group at higher velocity with respect to 
the main system, which includes the two subclusters A2440B and A2440C: these
two components must have recently crossed each other or 
are close to merging in the plane of the sky. 

We also applied the normality tests to the three subclusters. 
In the case of the $A$ subcluster, the 10 ROSTAT tests and Dip test exclude 
normality at more than the $10\%$ level. This result remains unchanged when 
excluding a low velocity galaxy that is 3000 km/s offset 
from the major velocity peak. We note however that the 
analysis of this subcluster relies on only 11 (10) velocities. 
For subclusters $B$ and $C$, all the normality tests (except for one in the case of $B$) and the
Dip test are consistent with a unimodal Gaussian distribution.

\subsection{Dynamical analysis}

\subsubsection{Mass estimates}\label{sec::mass_a2440}
 
We estimated the mass of the three components in A2440 following the same
method as for A2933. However, to determine the
harmonic radius we selected the galaxies in the red sequence,
defined using our $B$ and $R$ photometric catalogues, with $R < 20$;
in this way, background contamination was minimized. 
For each component, we selected galaxies satisfying the above criteria
within a radius of 3.0 arcmin from the respective BCG.

As a {\it caveat}, we emphasize that we consider subclusters 
potentially in a post--merger phase, in which case one expects important 
distortions in morphology and velocity field. 
However, previous analyses have shown that the velocity distributions 
in both components B and C can be assumed to be Gaussian.
We therefore assume that the two main components, A2440B
and A2440C, are now not far from dynamical equilibrium. 
In contrast, A2440A has a non--Gaussian velocity distribution, 
but this was determined using only a few redshifts. 
However, its characteristics are similar to those of a group, and its contribution to the whole mass 
should be {\em a priori} negligible. 

We find for A2440A, A2440B, and A2440C, respectively, that  
$M_{200} = 0.1 \pm 0.02 \times 10^{14}$ M$_\odot$, 
$M_{200} = 5.6 \pm 1.1 \times 10^{14}$ M$_\odot$, and
$M_{200} = 5.4 \pm 1.1 \times 10^{14}$ M$_\odot$.
As previously discussed, $B$ and $C$ appear to be the main 
subclusters; the estimated total mass of A2440 is
$M_{A2440} \sim 1.2 \times 10^{15}$ M$_\odot$.

\subsubsection {A2440: Two-body model}

The two-body model was  applied to the central system composed of A2440B 
and A2440C, with input parameters $R_p=0.442$ Mpc and $V_r=200$ km/s derived 
from the BCGs. As expected in terms of both  projected separation and 
velocities, the two-body model is not well constrained by the observations. 
Bound solutions exist  in both the pre-merger case with $t_0=12.2$ Gyr (two 
incoming and one outgoing), and in the post-merger case (one outgoing 
for $t_0<0.5$ Gyr, two incoming, and one outgoing for $t_0>0.5$ Gyr). 
Therefore, according to the two-body analysis, A2440B  
and A2440C may be systems seen either before or after first encounter.
 
\begin{figure}
\centering %Fichier absent
\includegraphics[angle=0,width=0.9\columnwidth,keepaspectratio] 
{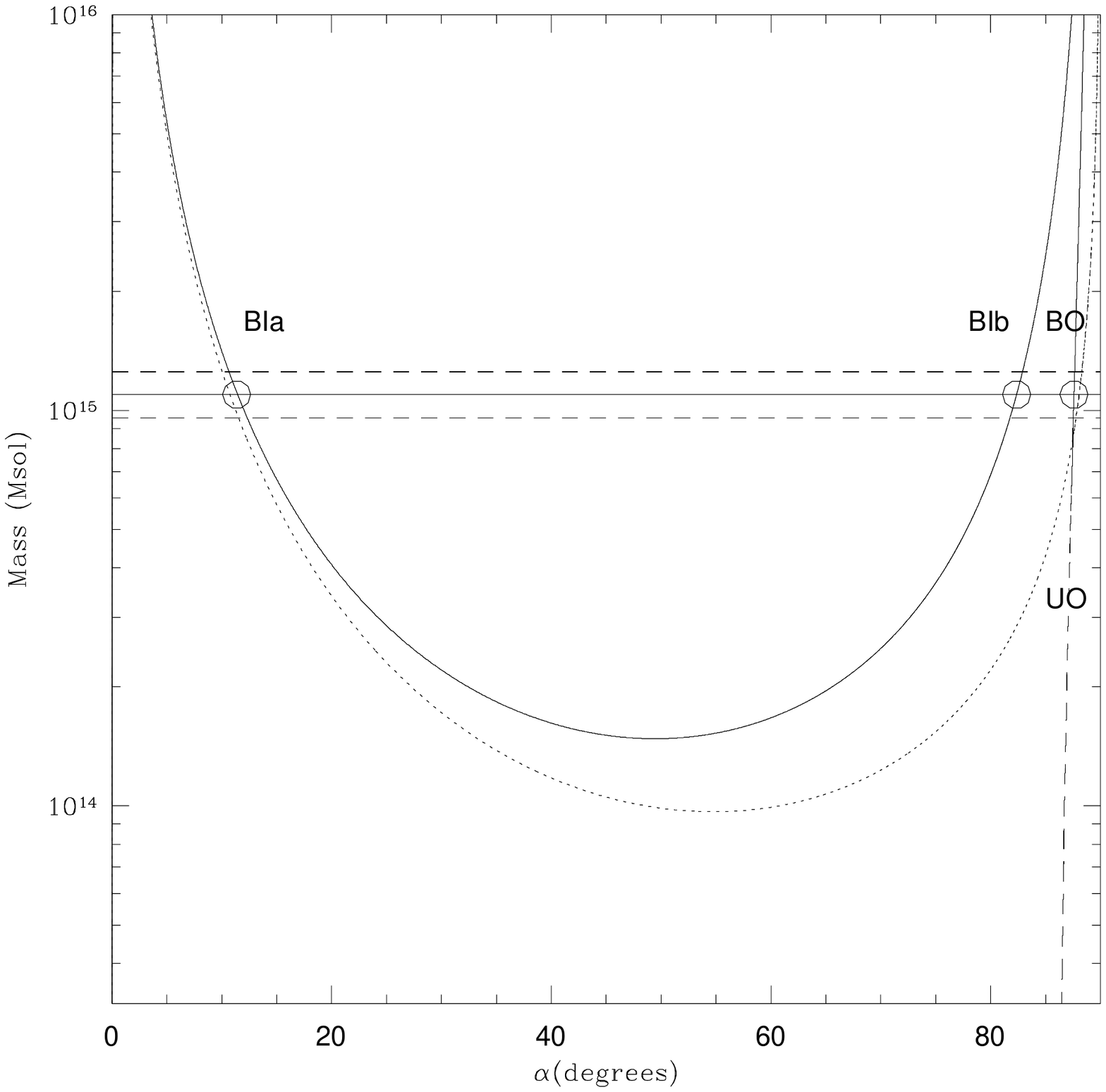}
\caption[]
{Two-body model for the system A2440A/A2440(B+C).  
The horizontal lines show the mass estimate (full line) and errors 
(dashed lines) (see section \ref{sec::mass_a2440}). 
The projected distance of 0.6 Mpc and the radial velocity difference at rest 
of 730 km/s were derived from the position and velocity of the
two centroids. The two systems are assumed to have been 
at zero separation 12.2 Gyr ago. 
The dotted line represents the Newtonian criterion for gravitational binding. 
Two bound incoming (BIa and BIb) and one unbound outgoing (UO) solutions
are found compatible with the mass estimates.}
\label{fig:A2440_2body}
\end{figure}

We  also tested the two-body system composed of A2440A and the whole 
complex A2440(B+C). To estimate $R_p$ and $V_r$, 
we used the position  and velocity of the centroid 
of the whole system A2440(B+C), obtaining $Rp=0.6$ Mpc and $V_r= 730$ km/s. 
In the pre-merger hypothesis ($t_0=12.2$ Gyr), there are two bound incoming 
(BIa, BIb), and one bound outgoing (UO) 
solutions (Fig. \ref{fig:A2440_2body}). If seen after first passage, one bound 
outgoing solution exists (BO). 
The different solutions are listed in Table \ref{table::2body}.

\subsection{The scenario for A2440}

We can now use the complete set of information collected from our analysis 
of A2440 to determine the merging scenario. The bimodal X-ray emissivity of 
the main component is evidence of the interaction of two massive subclusters 
being likely to undergo  or
to have undergone a merging event. 
At first sight, the existence of two units with cool cores, and a close
correspondence between the gas and bright galaxy distributions, which are well 
centered on the BCGs, are indicative of a pre-merger event. This was also 
the conclusion of Mohr et al. (1996) for A2440, who observed
the similarity between
galaxy projected density maps and X-ray emission maps from Einstein.

However the pre-merger hypothesis is strongly disfavored by the lack of any 
strong temperature enhancement between the two maxima, as we would expect
in a pre-merger event (RS01). The similar velocity 
distributions of the two subclusters can be understood in  terms of both a pre-merger and post-merger, but 
with different implications: in the pre-merger case it is indicative of
a merging in the plane of the sky, while in the post-merger case it is due to a very recent collision. 
The observed difference of clustering for faint and bright galaxies, with a 
good coincidence between the gas and bright galaxies centroids for A2440-B and 
A2440-C subclusters, but a segregation between the centroids of the gas and faint galaxies distributions also indicate that merging has already occurred. These
 results have 
already been found for several post-merger clusters (Biviano et al. 1996; 
Maurogordato et al. 2008). 

The evidence of a cold front delimiting the core of the Southern subcluster 
also favors a post-merger scenario, where the stripped subcluster is now 
moving away from its companion, as shown for instance in A1201 by 
Owers et al. (2009). The velocity offset of BG2 with respect to 
A2440-C also implies that there is some dynamical activity in this region.
To make this scenario compatible 
with the observed bimodal structure, we argue that the collision must 
have occurred with a relatively high impact parameter because a too small 
impact parameter would have led to the disruption of the two former subcluster
 cores during the violent relaxation of the gas. The similar values obtained 
for the temperatures and velocity dispersions of A2440B and A2440C suggests 
that these components have comparable masses. If one identifies the hot structures surrounding A2440B and A2440C as remnants of a shock wave, a scenario with
an equal mass 
collision with a high impact parameter after the maximum core collapse, when the 
shock wave begins to propagate towards the outskirts of the new 
structure, is therefore very likely to explain the A2440(B+C) complex.

The A2440A subcluster is probably a group, with a significant velocity offset 
($\sim 750$ km/s) with respect to the main cluster A2440(B+C). 
The coincidence of the BCG position with the maximum of the gas and galaxy 
distribution and the consistency of the BCG velocity with the subcluster 
mean velocity, suggest that the BCG is at the center of the subcluster 
in dynamical equilibrium. We should expect a Gaussian velocity distribution 
for this subcluster, which is not the case: this may be due 
to the small number of available redshifts (10) and  contamination
from galaxies not belonging to A2440A. 

The lack of signatures of gas compression between A2440A and A2440B excludes 
a pre-merger hypothesis implying small physical separations (BIa). The two viable 
pre-merger solutions are then incoming (BIb) or outgoing (BO), both with a 
major component along the line of sight ($\alpha \sim$ 80 degrees). 
The post-merger case remains unlikely, as there is a good 
correspondence between the gas and galaxy distributions and the X--ray
temperature map does not show any evidence of shocked gas in the periphery 
of A2440A. 
Therefore A2440A is probably a group infalling for the first time into the 
main cluster component along the NE/SW axis.

\section{A2384}

\subsection{X-ray gas morphology and thermal structures}

Looking at Fig.\ref{fig:OLT_a2384}, it is clear that 
the X-ray morphology of A2384 is 
very peculiar. The general shape 
of the whole cluster is very elongated, including a northern primary maximum, 
connected to a southern secondary maximum by a 
continuous gas distribution. The temperature of this gas is around 3-4 keV, 
while its surroundings are hotter (up to 4-5 keV). Both  subclusters 
exhibit cool cores. Another interesting feature
 in the temperature map is the 5 keV region 
spreading along the eastern region. 

\subsection{Galaxy density distribution }

A2384 is one of the best examples of colinear distribution of
substructures in a cluster (see West, Jones, Forman 1995). 
At all magnitude limits, the galaxy distribution shows a very 
elongated structure, extending over 1.2 Mpc along the N-S axis 
(Fig.\ref{isodens_mag}); embedded in this structure, there are 
two main subclusters roughly centered on two bright galaxies 
(which we  call BCG1 and BCG2 in the following; 
see Fig.\ref{A2384_spectro}). The 
northern subcluster is more densely populated than the southern one. At 
bright magnitudes (R$<$18), the galaxy distribution is strongly concentrated
in the northern (A2384N) component. 
Besides BCG1, in A2384N there are two other very bright 
galaxies, while in A2384S there are no bright galaxies around BCG2. 
%and Southern (A2384S) components. 
Near A2384N, a clump of galaxies to the west causes an elongation in 
the density map. While 
A2384S is very elongated along the general N/S axis of the system, 
the internal 
contours of A2384N are oriented towards the NW/SE, and rotate to 
the W-E and general NE/SW direction on larger scales.  
When including fainter objects, the isodensity contours of 
the two components are elongated towards each other (see Fig.\ref{isodens_mag}). 

\subsection{Galaxy velocity distribution }

%%% ADD
Before our observations, very few redshifts were available in the
literature for this cluster. In our 30' $\times$ 30' field, only 4
galaxies  already had a redshift; we reobserved two of them, and our
redshifts are in agreement with the literature values within the estimated errors.
Our final catalog includes 84 redshifts, and 
the number of cluster members is 56.
%%% END ADD

The mean location of the cluster,  CBI$=28263 \pm 154$~km/s, is well 
defined, but the scale is quite high (SBI$=1114\pm 120$~km/s).
The velocity distribution (Fig.\ref{fig:A2384_histo_emmix}) 
is very broad and has three main peaks. 
This extended velocity distribution could be either due to a very deep
potential (massive system) or multi-modality. The hypothesis of a massive 
system is improbable, given the low cluster optical luminosity and richness. 
The battery of normality tests do not indicate significant 
 departures from normality. The Dip test excludes unimodality with a P-value of 0.1.

\begin{figure*}
\centering
\includegraphics[angle=0,keepaspectratio,width=\columnwidth]{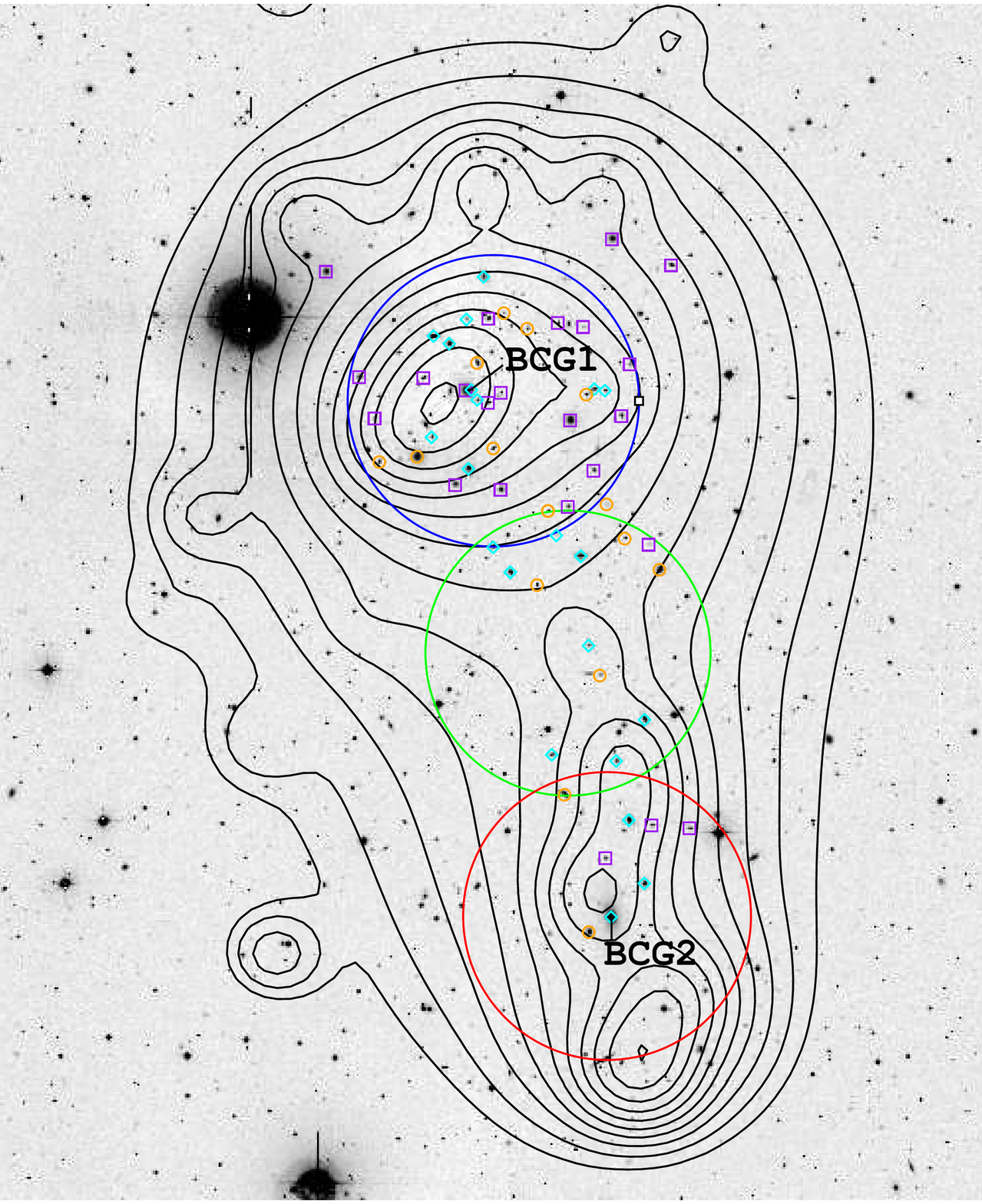}
\caption[]
{WFI R--band image (20'x25') centered on A2384.
Galaxies identified as cluster members from spectroscopy (flags $0$ and $1$) 
are marked with different symbols as in the previous cases 
(purple squares: [24000,28000] km/s, cyan diamonds: [28000, 29000] km/s and 
orange circles: [29000,31000] km/s). Isocontours corresponding to red sequence
galaxy density maps at $R<19$ are superimposed. The velocity histograms 
of the three regions delimited by circles with 3 arcmin radius, 
which include the two subclusters and the intermediary region, are displayed 
in Fig.\ref{fig:A2384_histo_subclus_3col}. North is up and East is to the 
left.}
\label{A2384_spectro}
\end{figure*}

\begin{figure}
\includegraphics[angle=0,width=0.9\columnwidth,keepaspectratio]{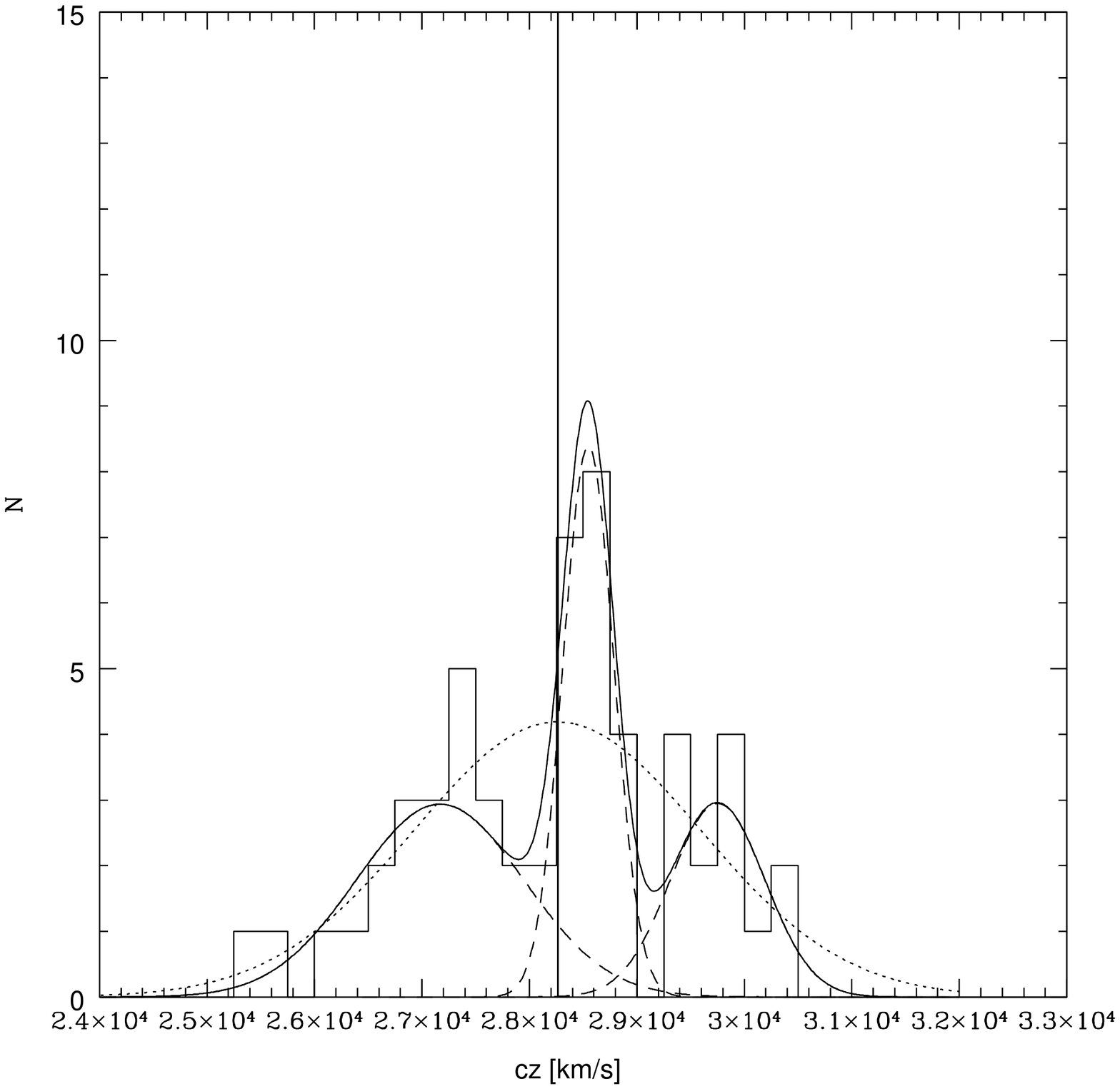}
\caption[]{Velocity histogram of A2384 (binning of 250 km/s). 
 The best Gaussian fit for the whole distribution
(dotted line) is centered on the vertical solid line which gives the location
value. Location and scale of the Gaussian were 
estimated with ROSTAT. We also show the three Gaussian functions (dashed lines)
corresponding to the best mixture model by EMMIX and the composite function 
(solid line).}
\label{fig:A2384_histo_emmix} 
\end{figure}

In Fig.\ref{A2384_spectro}, cluster members in the three velocity peaks are 
identified with different symbols and colors. 
We have very few redshifts in A2384S, because it is a poor structure. 
A striking feature is the high 
fraction of objects in the filament joining A2384N and A2384S, which
are associated to the second velocity peak ($\sim 28\,500$ km/s).

For A2384, the best--fit solution was found by EMMIX for a partition of 
three Gaussians roughly centered on the peaks visible in the velocity 
histogram, with a P-value of 0.16. This indicates that, while the fit 
quality is greatly enhanced by using 3 Gaussians instead of 1, the null 
hypothesis of 
Gaussianity is not rejected. The three Gaussian functions are visualized 
in Fig.\ref{fig:A2384_histo_emmix}; the sum of these functions 
reproduces the data quite well. The results are summarized in Table 
\ref{table::emmix}. The whole velocity distribution can be reproduced by the 
combination of two Gaussians with intermediate velocity distributions 
($\sim 365$ km/s and $\sim 665$ km/s corresponding to partitions 1 and 3, 
respectively) at mean velocities $\sim 1200$ km/s higher and 
lower than the mean velocity of the whole cluster, and by a third  strongly 
peaked component of low dispersion, $\sigma \sim 250$ km/s peaked at $\sim 28\,500$ km/s (corresponding to partition 2). 
In the following section, we discuss the spatial distribution of the galaxies 
assigned to the different velocity partitions with respect to the X-ray and 
optical density distribution.
 
\begin{figure}
\includegraphics[angle=0,width=0.9\columnwidth,keepaspectratio] {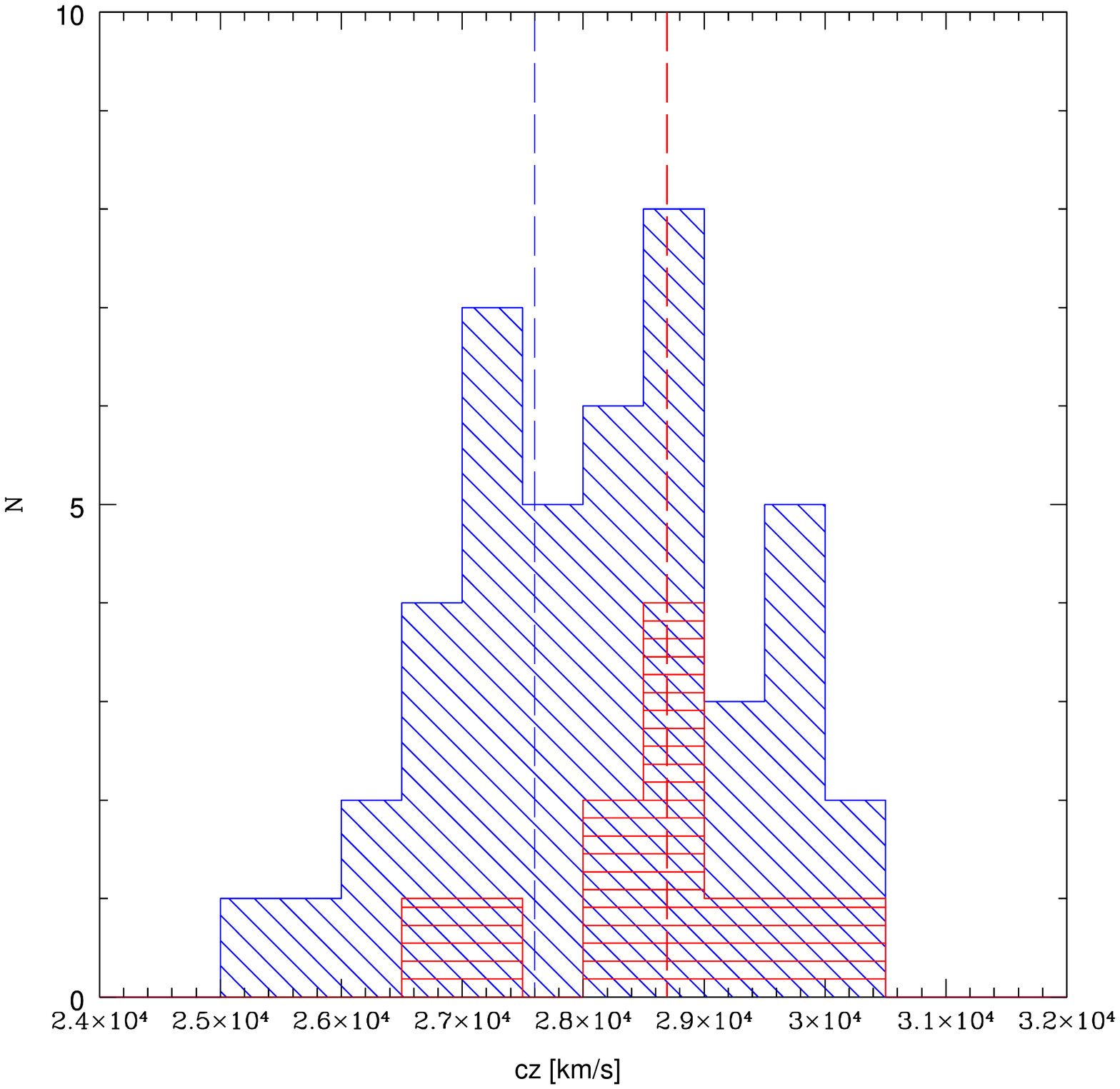}
\caption[]{Velocity histograms of A2384N (blue), and
A2384S (red). For each subcluster we included galaxies within a radius of 
5' from its center. Velocity bins of 500 km/s are used. 
Dashed lines: velocities of the brightest galaxies in A2384N (blue) and
A2384S (red).}
\label{fig:A2384_histo_subclus_2col} 
\end{figure}

\begin{figure}
\includegraphics[angle=0,width=0.9\columnwidth,keepaspectratio]{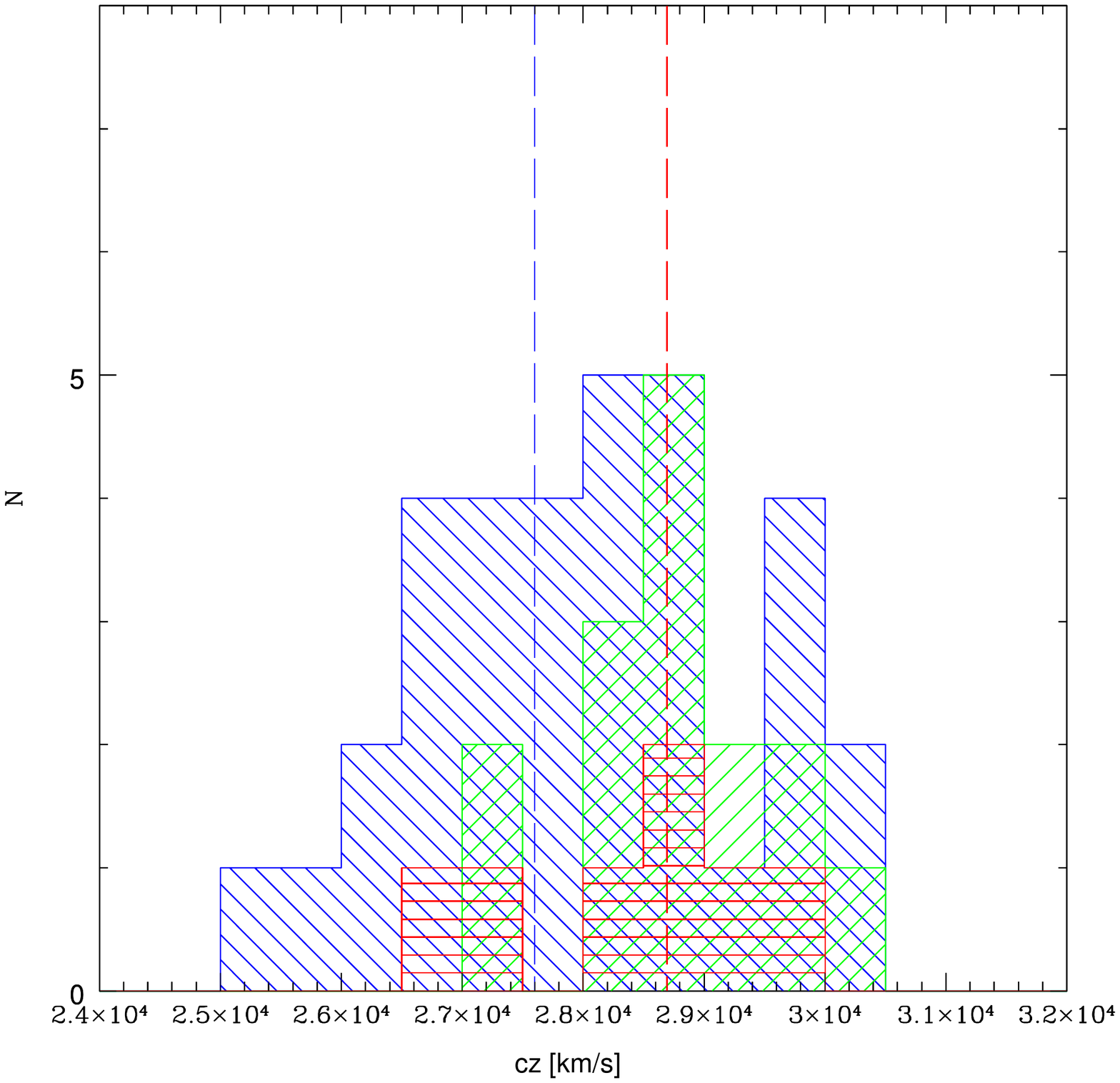}
\caption[]{Velocity histograms of A2384N (blue), the intermediate region 
(green) and A2384S (red). 
For each subcluster we included galaxies within a radius of 
3' from its center. Velocity bins of 500 km/s are used. 
Dashed lines: velocities of the brightest galaxies in A2384N (blue) and
A2384S (red).}
\label{fig:A2384_histo_subclus_3col} 
\end{figure}

\subsection{A2384: X-ray/optical combined analysis}\label{sec::a2384_OX}
Looking at Fig.\ref{fig:OLT_a2384}, both the 
gas and the galaxy maps approximately define the same bimodal structure. 
The northern and southern overdensities (A2384N and A2384S) 
correspond to the cool cores in the X-ray temperature map. 
%The location of the
% Brightest Cluster Galaxies with respect to the X-ray maxima of the two 
%subclusters has also been investigated. 
The X-ray maximum of A2384S is coincident with the BCG2 position, 
while BCG1 is slightly offset westwards (15 arcsec) with respect to the X-ray 
maximum of A2384N (see Fig.\ref{fig:OLT_a2384}). This offset cannot be due to
astrometric errors, as the precision of the optical images is superior to 1'', 
while the precision of X-ray images is superior to 4'', and there is a very good
correspondence between the optical and X-ray positions of point sources.
As BCG1 also has a significant velocity offset
with respect to A2384N (see below), this indicates that it is not at rest 
in the subcluster
potential. However, the velocity distribution of A2384N 
has a large dispersion, and at the cluster redshift 15 arcsec corresponds to 
a physical projected separation of about 25 kpc, which means that BCG1 
is close to the cluster center, which is probably included in its extended halo.

As already noted, the velocity dispersion of A2384 is unexpectedly high 
with respect to its optical luminosity and to the low values of its X-ray 
luminosity and temperature. A cluster with velocity dispersion 
$\sim 1000$ km/s should have an X-ray temperature $T \sim 7 KeV$, 
according to the $\sigma-T_X$ relation (Fig.\ref{fig::sigma_kt}), which is 
much higher than the observed value.
 Moreover, from a weak lensing analysis Cypriano et al. (2004) fit a 
velocity dispersion of $737 \pm 126$ km/s or $797 \pm 108$ km/s depending 
on the assumed profile (SIS and SIE, respectively). These velocity dispersions
corresponds to $T \sim 3.45$~keV and $T \sim 4.02$~keV, respectively, 
which are consistent
with the observed values in X-rays. 
This suggests that the measured velocity dispersion is overestimated.
%because of the presence of multiple velocity components.
 
We analyzed the velocity distribution in A2384N and A2384S, 
selecting galaxies in circles of radius 5 arcmin centered on the 
X--ray maxima. The velocity histograms in the corresponding regions are shown 
in Fig.\ref{fig:A2384_histo_subclus_2col}.
The mean velocity of A2384S is $\sim 675$ km/s higher 
than that of A2384N. The same trend is followed by the BCGs of the two 
subclusters, which show an even larger offset of $\sim 1000$ km/s. 
We also found that the position of the BCG1 has a small offset 
with respect to the centroid of 
A2384N, while the position of BCG2 corresponds to the centroid of A2384S. 
The velocity of BCG1 is 600 km/s lower than the mean velocity of
galaxies in A2384N, while the velocity of BCG2 is consistent with 
that of A2384S (see Table \ref{table::rostat}).

The velocity dispersions of the two subclusters are also quite large: 
$\sim 1200 $ km/s for A2384N and $\sim 900$ km/s for A2384S.
Both subclusters are above the $\sigma-T_X$  relation; 
the deviation is at the $3\sigma$ level for A2384N, 
but within the errors in the case of A2384S.
%(subclusters with $T \sim 3$ keV should have $\sigma \sim 650$  km/s). 
ROSTAT normality tests reject the Gaussian 
hypothesis for neither A2384-N, nor A2384S. The Dip-test however 
excludes unimodality for A2384-N 
(A2384-S) with a P-value of 0.1 (0.01). 
This suggests that the subclusters are still dynamically perturbed.
 
We also defined three subsamples by selecting galaxies within a 
radius of 3 arcmin from the centers of A2384N, A2384S, and the intermediate 
region (see Fig.\ref{fig:A2384_histo_subclus_3col}). 
The galaxies in the northern component show a large spread in their velocity
distribution, which is skewed towards lower values and has  
a peak at $\sim 28\,500$ km/s,
while galaxies in the southern and especially in the intermediate 
region are more strongly concentrated in the same peak; 
this peak corresponds to
EMMIX partition 2.

\subsection{Dynamical analysis}

\subsubsection{Mass estimates}\label{sec::mass_a2384}

We applied a  method similar to that applied to A2440,
determining the harmonic radius of each component by
selecting only the galaxies in the red sequence, with $R < 20$ 
and within a radius of 5 arcmin from the respective BCG. 

However, as we previously discussed, the velocity distribution
of this system is not simple. Our mass estimates could therefore be severely
affected. We find for A2384N and A2384S that 
$M_{200} = 1.4 \pm 0.3 \times 10^{15}$ M$_\odot$ and
$M_{200} = 0.6 \pm 0.1 \times 10^{15}$ M$_\odot$, respectively.
As the velocity dispersion is probably overestimated, we consider in 
the following $M_{200} = 2.0 \times 10^{15}$ M$_\odot$ as an upper bound to
the sum of the mass of the two components.

\subsubsection{Two-body model}

\begin{figure}
\centering
\includegraphics[angle=0,width=0.9\columnwidth,keepaspectratio]{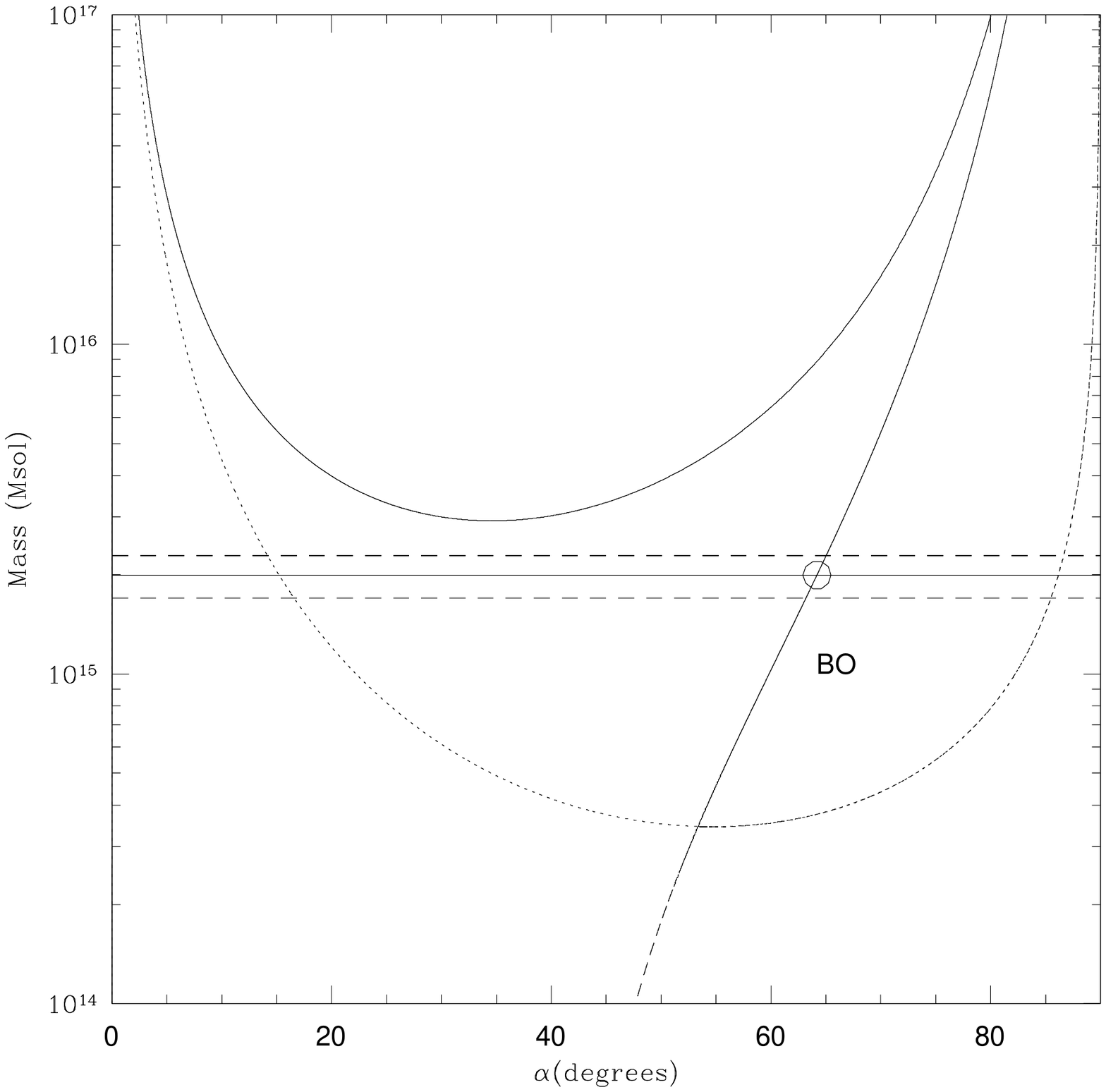}
\caption[]
{The sum of the virial masses of A2384N and A2384S 
as a function of the projection angle $\alpha$. 
The horizontal lines show the mass value 
(full line) and errors (dashed lines) as estimated in 
Sect.~\ref{sec::a2384_OX}. 
The projected distance of 1.14 Mpc and the radial velocity difference 
at rest of 1000 km/s were derived from the two BCGs positions and velocities. 
It is assumed that the two systems are 1.0 Gyr after the first passage. 
The dotted line represents the Newtonian criterion for gravitational binding. 
One unbound outgoing (UO) is found compatible with our mass estimates.}
\label{fig:A2384_2body}
\end{figure}

We applied the two-body model to A2384N and A2384S, 
assuming as values of $R_p$ and $V_r$ the projected separation and relative 
velocity at rest of the BCGs ($Rp= 1.14 Mpc $, $Vr= 1000$ km/s). 
We tested several scenarios: the pre-merger case ($t_0=12.2$ Gyr), 
and a range of post-mergers ($t_0$ = 0.2, 0.5, 1., 1.5 and 2 Gyr). In the pre-merger case, two incoming bound solution and one unbound solutions were found. In the post-merger case, only an unbound solution
 exists if the merging is recent ($t_0 < 0.5 Gyr$), and a bound outgoing 
solution exists if the merger is older ($t_0 > 0.5 Gyr$). 
Figure\ref{fig:A2384_2body} shows the case of a post-merger seen 1.0 Gyr 
after the encounter for which a unique bound outgoing solution exists. At $t_0 > 1.5 Gyr$, two bound incoming solutions and a bound outgoing solution are possible. 

\subsubsection{A2384 proposed scenario} 

The bimodal structure of A2384 detected in both the galaxy and gas density 
maps suggests that the two components either are about to or have undergone a merger 
event. The presence of two well--separated components with cold cores, 
coincident 
in the X-ray and  the optical, is often observed in the case of pre-mergers.

However, the trail of gas and the bridge of galaxies extending from 
A2384N to A2384S are difficult to explain if the subclusters have not yet 
interacted. 
If A2384 were a pre-merger, its temperature map should exhibit a compression 
region between the two subclusters, which is not the case. 
The velocity distribution shows a large dispersion ($\sigma \sim 1200$ km/s), 
inconsistent with the value of the temperature if it simply reflects the
mass of the system. This high value of the velocity dispersion 
and the velocity distributions 
of the two components being mixed imply that the merging has already occurred. 
A2384N shows several signatures of previous dynamical 
activity: a slight segregation between gas and galaxies, an offset of the 
position of the BCG from to the peak of the X-ray emitting gas, 
and of its velocity from  the mean velocity of the subcluster. 
This suggests that the dynamical state of A2384N is still disturbed possibly due to a previous collision with A2384S 
or/and a secondary merger event within A2384N itself.

At first sight, A2384S is more regular than A2384N. It has a good spatial coincidence between its gas
and galaxy distributions, coincidence between the position and velocity of 
the BCG and the X-ray center and mean velocity of the subcluster, respectively.
Its velocity distribution is very broad and  non--Gaussian, but 
this may be due to contamination by objects in the periphery of A2384N. 

This complex structure could be understood in a scenario 
in which a low mass cluster of galaxies has crossed the environment of a more massive one  and has been stripped of a large fraction of its gas and galaxies. 
The ``head'' of the system would then correspond to A2384S, and the tail of 
galaxies belonging to the original cluster affected by A2384N would lie 
in the intermediary region. 

We conclude that the most likely scenario for A2384 is a post-merger between 
two unequal mass clusters. However, if the merging has already occurred, 
we have to explain why the two cool cores have remained unaffected by 
the merging process. Numerical simulations (Poole et al. 2006) have shown that 
the initial cool cores 
of the primary and secondary components survive the first core crossing, 
and disappear after the second pericentric passage. 

It is however difficult to constrain the parameters
of the scenario, as the mass of A2384 is not well determined. 
From the 2-body analysis, bound solutions exists only if the merger event 
is older than 0.5 Gyr. Older mergers (1-1.5 Gyr) with larger physical 
separations between the units are favoured by the lack of signatures of 
compression in the X-ray gas (Table \ref{table::2body}). For mergers older 
than 1.5 Gyr, one can see that incoming solutions, corresponding to second 
infall, are again possible. However, these also predict small values of the real 
separation between the two components, which are unlikely from the lack of 
signature of compression of the gas in the temperature maps, as noted before. 
One is then left with the most probable hypothesis of two bound outgoing 
subclusters seen more than 1.0 Gyr after the first passage, with a collision 
near the line of sight. Figure \ref{fig:A2384_histo_subclus_2col} shows as an example the bound outgoing 
solution in a 1.0 Gyr post-merger, which is in good agreement with both 
our X-ray and optical observations.

\section{``Idealized'' simulation \& XMM emulation}

In the previous sections, we combined optical and X-ray observations 
to constrain the merging scenario for each one of our three bimodal systems. 
To help define the scenario and test its consistency, we 
tried to reproduce the observed X-ray properties with 
numerical simulations.
Adopting an approach similar to RS01, 
we placed two clusters in a box and allowed them to evolve under gravity 
in an adiabatic framework. The initial parameters of the 
collision are inferred from the X-ray and optical observational constraints. 

To this aim, we developed a set of programs that allow us to reproduce
the observational process and analyze the
simulated clusters in the same way as we analyze real ones, adopting
an approach similar to e.g. Gardini et al. (2004) for XMM observations.

Using an emission plasma model, where gas pressure and density are
known everywhere, we can compute the emitted X-ray photons as a function
of telescope area and exposure time. 
Fixing the direction of observation, we collect photons from the 
different cells along the line of sight, taking into account that we observe clusters in projection (as long as we can assume that the thin plasma approach is valid).

The first step in developing a good "idealized" merger simulation is to check the 
way we implement the hydrostatic equilibrium in each (sub)cluster (hereafter
``unit'') to be sure that, 
during collision, what we see is really due to the collision and not a
spurious result of numerical errors in the initial units. 
Once this is achieved (by allowing a 
unit to evolve alone in the middle of the box for a very long time), 
we are able to build collision simulations. 

The total mass profile of our "perfect cluster" follows a 
classic NFW profile, $\rho_{dm}(r) \propto
\frac{1}{r(1+r/r_S)^2}$ (Navarro et al. 1996), while the gas follows 
a profile 
described in Suto et al. (1998) adapted to keep the gas mass fraction
constant at large radii. The dark matter profile is
derived from the difference between the total mass profile and the gas. 
We note that the NFW profile is derived from  pure dark matter simulations, thus
this approach is likely to be safe as long as the gas mass fraction is not
too high (thus not in the central region of cooling-flow clusters).

The velocity dispersion of dark matter particles is computed (as
in RS01) using the virial equation
$\frac{d}{dr}[\rho_{dm}(r)\sigma^2(r)]=-\frac{GM(r)}{r^2}\rho_{dm}(r)$.

After fixing the gas-to-mass  ratio, the concentration parameter 
$c_{200} = \frac{M_{200}}{r_S}$, the 
mass ratio between units and the impact parameter 
(defined as the minimal distance between the two initial
trajectories in unit of the NFW scaling radius $r_S$), 
we reconstruct a pre-collision scenario by computing two units 
as described above. We place these two units in a rather large box 
to avoid any border effects. 
We compute the initial velocity considering the pre-merger
evolution as a free fall of the two units, exactly in the way
described in RS01. We then compute the evolution of the
system with the AMR hydro-NBody simulation RAMSES (Teyssier et al. 2002). At
each time step, for all the cells in the simulated volume, we store the
pressure, velocity, and density of the gas, and position and velocity of
dark matter particles. After running these simulations, 
the following steps are the computation of the photons 
emitted by the simulated volume and
their ray-tracing using the EPIC/XMM instrumental response. These steps are
fully described in Bourdin et al. (2004). We finally have a dataset 
that can be reduced in the same way as real XMM/EPIC observations.

All these previously described steps are important for a correct
comparison between observation and simulation. Because of the complexity of this procedure, it is impossible to explore the whole parameter space
when determining the optimal collision parameters. By 
looking simultaneously at the brightness, temperature maps of the X-ray gas,  
and galaxy density 
maps, we first guess a set of possible parameters that we place into the 
complete simulation stream and iterate manually to optimize the set of
parameters. Therefore, we do not pretend to fit the data, 
but we attempt to obtain a scenario reproducing
the main features of the observed merging system.

% Pre-Merger plus serre que A1750. MerPol37 (b=0 M/M=1).
\begin{figure}
\centering
\includegraphics[scale=0.45,angle=0,keepaspectratio,width=\columnwidth]
{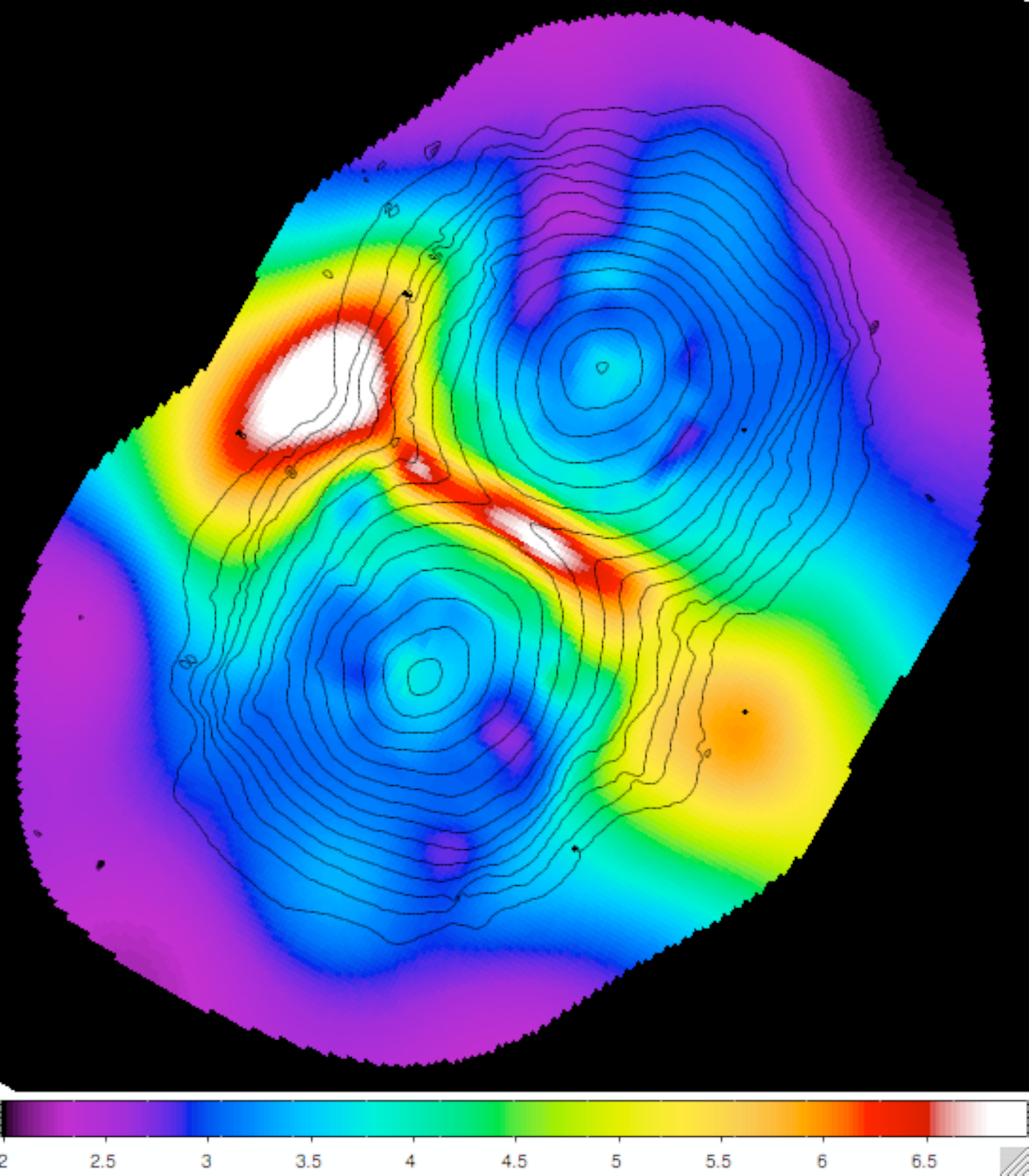}
\caption{Simulated wavelet reconstructed temperature map with luminosity 
contours superimposed, derived from the simulation which best reproduces 
the observed features of A2933.}
\label{fig:TLX_MP37}
\end{figure}
% MerPol37 equal mass regarder qd on a la "bon" contraste en T (Tmax:Tmean) 
% dans la barre de compression et en deduire une estim de l'age de A2933...

\begin{figure}
\centering
\includegraphics[scale=0.45,angle=0,keepaspectratio,width=\columnwidth]
{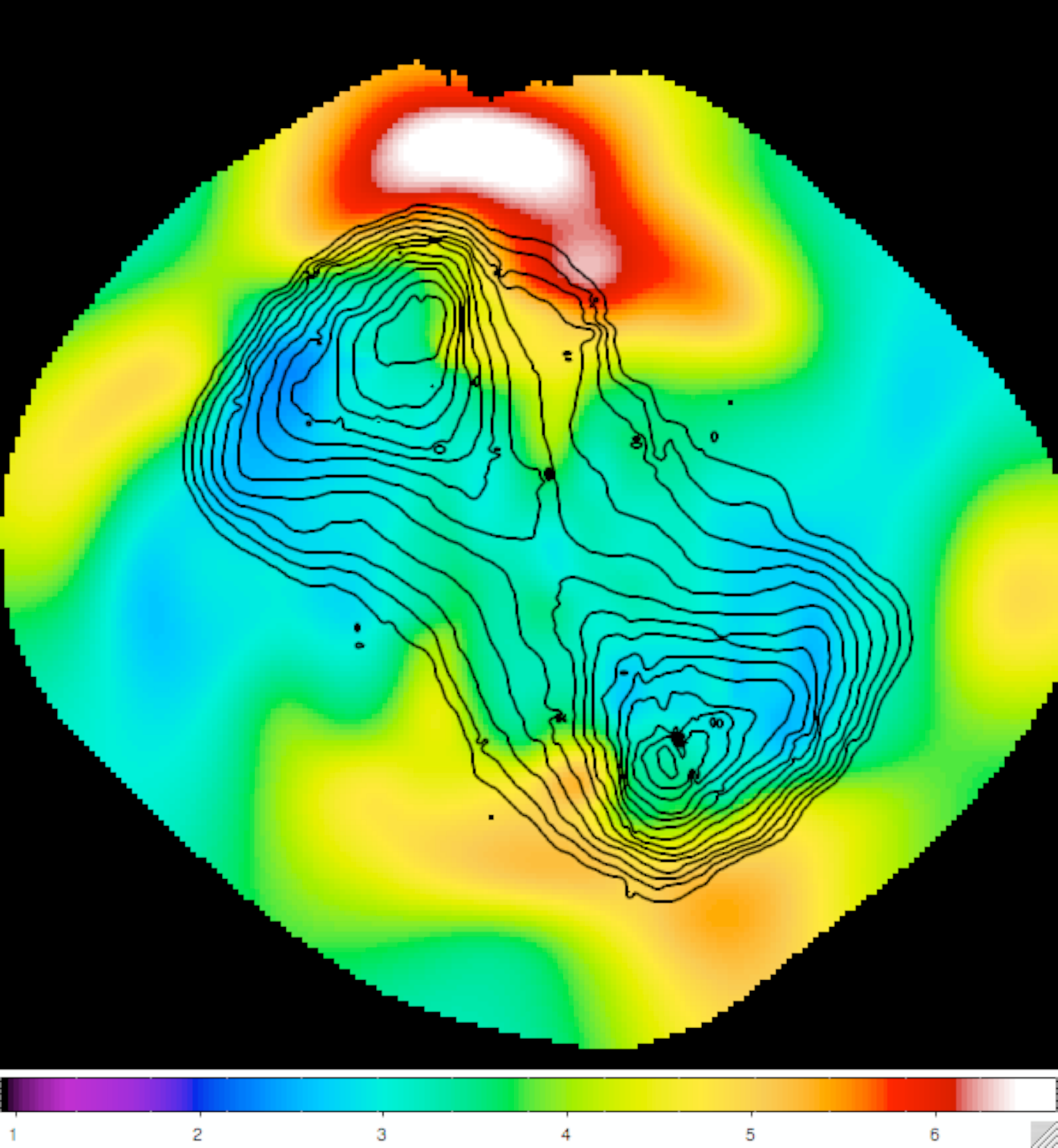}
\caption{Simulated wavelet reconstructed temperature map with luminosity 
contours superimposed, derived from the simulation which best reproduces 
the observed features of A2440.}
\label{fig:TLX_MP136}
\end{figure}

\section{Comparison with simulations and discussion}

Taking as input the scenario inferred from observations for the main merger 
event, we 
obtained density and temperature maps as they would be observed with XMM, 
and compared them with the real observations. 

In the case of A2933, the most likely scenario is a pre-merger of two 
components of comparable mass for which there is 
a small angle between the collision axis and the plane of the sky. 
We simulated an equal--mass merger in the plane of the sky with an impact 
parameter set to zero, and allowed it to evolve with time. The resulting temperature 
map and luminosity contours for this scenario are presented in Fig.\ref{fig:TLX_MP37}. The X-ray properties, in particular the temperature 
contrast between the high temperature region and the mean temperature of each 
unit and its relative angular distance, are closely reproduced 
by a pre-merger scenario $\sim 200 Myr$ before the core collapse. 

For A2440, which is expected to be a post-merger case, we simulated a variety 
of equal mass mergers with different impact parameters and tried to find the 
best configuration reproducing simultaneously the observed temperature and 
luminosity distributions. In Fig.\ref{fig:TLX_MP136}, we present our best 
candidate. This scenario  corresponds to a collision with an impact parameter 
of 15 $r_s$ seen 450 Myr after the maximum core collapse. We managed to 
reproduce the most striking features revealed by the observations, in 
particular the hot region surrounding the X-ray maximum of 
each unit, and the temperature contrast between the hottest region and the colder ones. Since our code follows a purely adiabatic approach, we do not expect of course to 
reconstruct the two cool cores.
%New Version pour A2384

In the case of A2384, we performed different trials allowing the 
impact parameter value to vary.
If the impact parameter is not zero, 
it should be large enough to allow the gas of the smaller unit to follow the influence of its original  potential well after the collision 
(we note that this is also necessary to explain the remaining "cool" core of 
each unit). 
At the same time, the impact parameter should  be small enough  to enable the gas to be stripped efficiently. In these limits, the gas is stripped 
along a rather significant distance as observed. Simulations by Poole et 
al. (2006) show a gas surface density similar to that of A2384, in the case of 
a 3:1 mass ratio seen at the first apocentric passage of the secondary 
cluster, at $\sim 1.5$ Gyr after the closest approach (see their Fig.4 central
 panel). In these scenarios, it is nevertheless very difficult to keep the 
axial symmetry observed in luminosity and in temperature maps and impossible 
to explain that the BCG of the south unit still lies at the X-ray maximum.

On the other hand, in a null impact parameter collision, if a small group 
encounters a heavier unit, its gas can escape significantly after the first 
maximum core collapse only if its central density is very high compare to the 
central density of the massive object. Simulations by Poole et al. (2006) 
are able 
to strip gas even with a null parameter thanks to their initial gas profile 
($S \propto r^{1.1}$) profile, which leads to a high inner gas density 
in the core.
In contrast, if the small unit has a concentration parameter higher than the 
massive one (say by a factor 2) then the gas appears sticked to its own unit 
even after the first maximum core collapse. Unfortunately, we cannot simulate 
such a complex scenario in the adiabatic scheme since we have no cooling thus 
the inner temperature of the small unit increases strongly with the 
concentration parameter, preventing the correct description of the gas 
luminosity and temperature during the collision.

\section{Conclusion}

We have performed a combined X-ray/optical analysis of 
three bimodal clusters at low redshift, 
selected as merger candidates from the sample of 
Kolokotronis et al. (2001). We have confirmed the disturbed dynamical state 
of  these clusters and reconstructed the general trends of their main merging 
scenario. 
A2933 has been shown to most likely be an equal--mass advanced pre-merger 
($\sim 200 Myr$ before the core collapse). A2440 is a recent equal--mass merger 
($\sim 450$ Myr after core collapse) with another subcluster infalling along 
the main axis of the cluster, and A2384 is the result of an older collision 
between two units of mass ratio 1:3 ($\sim 1.0-1.5$ Gyr after core collapse).

We emphasize the complementarity of the X-ray and optical data in determining
the parameters of the scenario (angle, mass-ratios, epoch) that were refined by running simulations of idealized cluster collisions. 
Our  analysis confirms the efficiency of 
the selection based on the comparison of gas and density maps 
to identify merging clusters when (as is frequently the case)
information on temperature and velocity distribution is not available. 
Moreover, the level of segregation between galaxies and gas provides some 
indication of the merging stage. This approach is far more powerful 
than  the density maps  taken individually in either X--rays or the optical. 
From X-ray density maps only, the three bimodal clusters presented
in this paper would have been expected to be in the very last phase before 
the first maximum core collapse, while two of them are shown to be 
post-mergers with surviving cores in the two units. 
The properties of the temperature maps  are essential 
to assess the merging stage: the hot region in between the two sub-clusters 
in A2933 is indicative of a pre-merger, the spectacular hot structures in the 
periphery of  A2440 and A2384 have been identified as remnants of a post-merger shock wave, and  the cold front detected in the S/W region of A2440 is 
consistent with a  a post-merger. On the other hand,  redshift information is needed to understand the dynamics of the system, 
determine the mass of the subclusters and the relative motion of the BCGs, and 
finally date the merger event and constrain its axis using  the two-body analysis.

Numerical simulations have allowed us to refine the collision parameters. 
Thanks to our XMM simulation tools, 
we correctly handle all projection and instrumental effects. 
Thus we can rely on these "observed simulations" and select the most likely scenario 
based on measured criteria such as distances, temperature, and luminosity.
We have been able to reproduce the main features of A2933 and A2440, 
simulating the collision of two systems initially in complete hydrostatic 
equilibrium (with an approach similar to RS01) and adding a complete set of 
programs to reproduce X-ray observations for simulations and obtain
 EPIC/XMM synthetic observations (with an approach similar to Gardini et al.
2004). The case of A2384 is too complex to be
described with our approach, but we have been able to place constraints on 
its collision parameters.

%{\bf The analysis of 
%the velocity distribution has also to be done in the light of other data.
%A ``blind'' application of normality tests do not exclude significantly the gau
%ssian hypothesis in most cases. 
%Sophisticated algorithms, fitting a mixture of 
%gaussians, bring a more complete description of the velocity distribution, but 
%do not exclude the gaussian hypothesis for the velocity distribution of the pos
%t-merging clusters A2440 and A2384, while bi-modality is clearly found in the c
%ase of the pre-merger A2933.}

The results obtained from the analysis of these clusters can be combined to 
our previous work on other MUSIC clusters
to draw a more general picture of merging clusters properties. 
In the three clusters studied in the present paper,
we have shown the existence of preferential axes following the general 
position angle of the cluster, as previously shown in other merging clusters 
(Arnaud et al. 2000; Plionis et al. 2003; Maurogordato et al. 2006). 
Subclusters generally host one or more BCGs. 
In most cases, the different merging subclusters and their brightest 
BCGs are aligned along this direction. This corroborates the property of 
preferential alignment of BCGs with their host cluster 
(Niederste-Ostholt et al. 2010) on the scale of subclusters. 
In A2384, we have detected a spectacular filament of galaxies and gas 
probably stripped from the colliding group along the merging axis.  

Another interesting aspect is the position and motion of the BCGs 
relative to their subclusters. We have found that
their angular coordinates generally correspond (within the errors) 
to the X-ray centroids of their host subclusters (A2933S, A2440A, A2440B, 
A2440C, A2384S) but in some cases there is a small offset (A2933N, A2384N).  
Similarly, in some cases the radial velocities of the BCGs are consistent with the mean 
velocity of their respective subcluster (A2440A, A2440B, A2384S), but in other cases  they are offset by 300-500 km/s (A2933N, A2933S, A2440C, A2384N). 
These offsets are signatures of ongoing dynamical activity in the 
subclusters, as found in A3921B (Ferrari et al. 2005) and in A2163A 
(Maurogordato et al. 2008), which were both identified as recent mergers. 
In an analogous way, the beginning of interaction between A2933N and A2933S, 
the probable recent mergers occuring within A2933N and A2384N, and the recent 
merger between A2440B and A2440C can explain the observed BCG offsets.

We have also found that, in addition to the main merging event, 
a large fraction of MUSIC clusters appear to contain secondary merging 
events  (A2933N and A2384N) and infalling groups,e.g., A2440 (this work), 
A521 (Ferrari et al. 2003), and A2163 (Maurogordato et al. 2008).

These results are consistent with a hierarchical scenario of structure 
formation where clusters form by successive mergers and accretion of matter 
along large-scale filaments, which may cause alignments with structures 
on various scales (Basilakos et al. 2006; Lee and Evrard 2007; 
Faltenbacher et al. 2008). 

Moreover, we have also shown that in recent post-merger clusters, 
such as A2440 (this work) and A2163 (Maurogordato et al. 2008), 
galaxies exhibit a strong luminosity segregation, similar to the
case of the Coma cluster studied by Biviano et al. (1996). 
We have also found a deviation from the $\sigma-T_X$ relation 
for the subclusters in a post-merger stage, the largest one 
(at the 3$\sigma$ level) being detected for A2384N. 
At variance, both pre-merging subclusters A2933N and A2933S follow 
the $\sigma-T_X$ relation.

The impact of the merging process on galaxy properties in these clusters, 
in particular on star formation, will be addressed in a forthcoming paper. 

To test the validity of these properties on a large sample, 
we are currently extending this work to a subsample of 
the C4 SDSS cluster sample (Miller et al. 2004) with both optical and X-ray 
available data.

\begin{acknowledgements}
The authors want to thank Romain Teyssier for the use of RAMSES Hydro-NBody 
code and Albert Bijaoui and Eric Slezak for providing their program 
estimating density maps through a multi-scale approach, and Monique Arnaud for fruitful discussions.
We thank the Programme National de Cosmologie et Galaxies  of CNRS for his 
constant support on this program, the Observatoire de la C\^ote d'Azur and the
Laboratoire Cassiop\'ee, CNRS, for specific funding of this project.
H.B. acknowledge financial support from contract ASI-INAF I/088/06/0.
We also want to thanks an anonymous referee for his/her comments and 
suggestions which helped us to improve the quality of the paper.

\end{acknowledgements}

\begin{longtable}{ccccccc}
\caption{Radial velocity measurements
in the field of Abell 2933.
   The full catalogue  is
   available in electronic form at {\rm www.edpsciences.org}} \\
\label{Table_A2933}\\
\hline
\hline
Galaxy & Right Ascension(J2000)  & Declination (J2000)  & v (km~$s^{-1}$) & 
$\epsilon$ (km~$s^{-1}$) & $R_{TR}$ & Flag\\
\hline
(1)  & (2) & (3) & (4) & (5) & (6) & (7)\\
\hline
   1 & 1~40~14.73 &- 54~32~36.2 &   184 & 19 &16.8 & 0\\
   2 & 1~40~16.47 &- 54~29~51.7 & 26159 &104 & 2.7 & 0\\
   3 & 1~40~17.54 &- 54~28~15.3 & 27500 &121 & 3.3 & 0\\
   4 & 1~40~18.80 &- 54~28~45.6 & 27303 & 80 & 4.3 & 0\\
   5 & 1~40~20.24 &- 54~29~13.4 & 26356 &101 & 3.8 & 0\\
   6 & 1~40~22.09 &- 54~29~28.6 &   152 & 50 & 5.6 & 0\\
   7 & 1~40~22.15 &- 54~29~28.6 &   137 & 21 &16.0 & 0\\
   8 & 1~40~22.98 &- 54~32~10.6 & 26623 & 33 &10.5 & 0\\
   9 & 1~40~25.52 &- 54~28~54.4 & 26489 & 66 & 4.2 & 0\\
  10 & 1~40~25.84 &- 54~30~49.4 & 26018 &106 & 3.0 & 0\\

\hline
\end{longtable}

\begin{longtable}{ccccccc}
\caption{Radial velocity measurements
in the field of Abell 2440.
   The full catalogue  is
   available in electronic form at {\rm www.edpsciences.org}} \\
\label{Table_A2440}\\
\hline
\hline
Galaxy & Right Ascension(J2000)  & Declination (J2000)  & v (km~$s^{-1}$) & 
$\epsilon$ (km~$s^{-1}$) & $R_{TR}$ & Flag\\
\hline
(1)  & (2) & (3) & (4) & (5) & (6) & (7)\\
\hline
   1 &22~23~21.33 &-  1~40~54.0 & 58154 & 54 & 3.5 & 1\\
   2 &22~23~22.01 &-  1~42~31.2 &  -136 & 13 &18.4 & 0\\
   3 &22~23~23.56 &-  1~42~19.5 & 27293 & 27 & 7.2 & 0\\
   4 &22~23~24.82 &-  1~41~38.1 &  -134 & 57 & 4.7 & 1\\
   5 &22~23~27.64 &-  1~40~57.3 & 26389 & 14 & 0.0 & 0\\
   6 &22~23~28.80 &-  1~42~28.6 &   -51 & 11 &23.0 & 0\\
   7 &22~23~30.06 &-  1~41~50.0 & 26670 & 19 &15.1 & 0\\
   8 &22~23~30.75 &-  1~40~44.5 & 92248 & 41 & 5.6 & 0\\
   9 &22~23~31.92 &-  1~42~21.2 & 27415 & 22 &13.4 & 0\\
  10 &22~23~33.70 &-  1~39~58.8 & 26784 & 54 & 5.2 & 0\\

\hline
\end{longtable}

\begin{longtable}{ccccccc}
\caption{Radial velocity measurements
in the field of Abell 2384.
   The full catalogue  is
   available in electronic form at {\rm www.edpsciences.org}} \\
\label{Table_A2384}\\
\hline
\hline
Galaxy & Right Ascension(J2000)  & Declination (J2000)  & v (km~$s^{-1}$) & 
$\epsilon$ (km~$s^{-1}$) & $R_{TR}$ & Flag\\
\hline
(1)  & (2) & (3) & (4) & (5) & (6) & (7)\\
\hline
   1 &21~52~ 2.86 &- 19~41~36.4 & 26810 & 35 & 7.6 & 0\\
   2 &21~52~ 3.13 &- 19~36~57.7 &    18 & 41 & 8.6 & 0\\
   3 &21~52~ 4.20 &- 19~39~20.8 &   -69 & 23 &16.3 & 0\\
   4 &21~52~ 4.45 &- 19~30~17.7 & 27250 &120 & 0.0 & 1\\
   5 &21~52~ 5.01 &- 19~32~45.3 &   476 & 94 & 3.5 & 0\\
   6 &21~52~ 5.44 &- 19~36~24.9 & 29289 & 57 & 6.2 & 0\\
   7 &21~52~ 6.10 &- 19~41~33.0 & 27472 & 25 &10.0 & 0\\
   8 &21~52~ 6.36 &- 19~35~54.6 & 27454 & 67 & 4.5 & 0\\
   9 &21~52~ 6.71 &- 19~39~25.7 & 28590 & 88 & 3.3 & 0\\
  10 &21~52~ 6.72 &- 19~42~43.1 & 28927 & 85 & 2.8 & 1\\

\hline
\end{longtable}

\end{document}